\documentclass[a4paper,11pt]{article}
\pdfoutput=1 
\usepackage{jheppub} 

\usepackage[T1]{fontenc} 
\usepackage{appendix}
\usepackage{amsmath}
\usepackage{braket}
\usepackage{indentfirst}
\usepackage{comment}
\title{\boldmath Modeling Compact Objects with Effective Field Theory I: The Effective Action }

\author[]{Irvin Martinez}

\affiliation[]{High Energy Physics, Cosmology \& Astrophysics Theory Group, Department of Mathematics \& Applied Mathematics, University of Cape Town, Cape Town, 7701, South Africa}

\emailAdd{mrtirv001@myuct.ac.za}

\abstract{This is part 1 of 3 from the master's thesis: Modeling Compact Objects with Effective Field Theory, supervised by Amanda Weltman. Using the Effective Field Theory framework for extended objects and the coset construction, we build the leading order effective action for the most general compact object allowed in an effective theory of gravity as general relativity. By recognizing the symmetry breaking pattern of a charged spinning compact object, we derive all the covariant building blocks and constraints to build up the relevant invariant operators in the action to all orders. We derive the effective action for a spinning extended object and make the connection to currently used theories, which use conjugates variables. Moreover, we build the action of a charged particle as well, and include the size structure effects, such as tidal deformation, polarization and dissipation in the quasi-static limit. The invariant operators in the action are accompanied by coefficients that encapsulates the properties of the compact object. We match the known coefficients of the effective action from the literature, and point out the unknown ones that are to be derived. 

}

\begin{document} 
\maketitle
\flushbottom

\section{Introduction}
\label{sec:intro}

The modeling of compact objects and their interactions must take into account different effects, such as the spin, charge, and the stellar structure of the various stellar types. In this work we bring the tools of effective field theory (EFT) for extended objects \cite{Goldberger:2004jt, Goldberger:2005cd, Delacretaz:2014oxa} to model BHs and NSs as point particles, with additional effects encoded as higher order corrections in the action. 

Although at first it might seem counter intuitive to model such massive objects as point particles, there is in fact, a lot of similarity in the description of compact objects and elementary particles \cite{Arkani-Hamed:2019ymq, Moynihan:2019bor}. For instance, the no hair theorem implies that a BH effectively behaves as a point particle, and it has been shown that, by matching on to the effective action of a spinning point particle, a spinning BH is described relativistically to all orders in the multipole expansion from minimal coupling in the amplitudes formalism \cite{Guevara:2018wpp, Chung:2018kqs}.

Using the coset construction \cite{Coleman:1969, Callan:1969sn,Volkov:1973vd, Ivanov:1975zq} in the Effective Field Theory (EFT) framework for extended objects \cite{Goldberger:2004jt, Goldberger:2005cd, Delacretaz:2014oxa}, we build the most general effective action allowed in an effective theory of gravity such as General Relativity with electrodynamics. In this theory, BHs can be described by only three parameters, its mass, spin and charge, behaving effectively as a point particle \cite{Arkani-Hamed:2019ymq,Moynihan:2019bor}. One of the advantages of using the EFT for extended objects is that we can model the compact objects as point particles, with its additional effects and internal structure encoded as higher order corrections in the action, which are made up of the allowed invariant operators given the symmetries of the objects. These operators are accompanied by the coefficients of the effective theory, which encapsulates the internal properties of the compact objects. In this sense, we construct the lowest order effective action for the most general allowed compact object in the point particle approximation, which is one that is charged and spinning.

The coset construction is a very general technique from the EFT framework that can be used whenever there is a symmetry breaking. With this technique, it is possible to derive the covariant building blocks that transform correctly under the relevant symmetries, which then can be used to form invariant operators to build up an effective action. Any state other than vacuum will break some of the space-time symmetries, and by correctly identifying the pattern of the symmetry breaking, we can derive its effective action. Within this approach, we can treat the coefficients that appears in front of the invariant operators as free parameters to be fixed by experiments or observations.  In this sense, we are interested to know what was the full symmetry group, G, of the EFT, and what subgroup, H, was realized non-linearly, parametrized by the coset, G/H \cite{Penco:2020kvy}.

An effective theory of gravity, such as General Relativity, can be derived using the coset construction by weakly gauging the space-time symmetry group of gravity, the Poincaré group ISO(3,1), and realizing translations nonlinearly, which is parametrized by the coset ISO(3,1)/SO(3,1) \cite{Delacretaz:2014oxa}, with SO(3,1) the Lorentz group. From this coset parametrization, it is possible to derive the widely known Einstein's vierbein theory of curved space or tetrad formalism, which is a generalization of the theory of General Relativity that is independent of a coordinate frame. Once the underlying theory of gravity is developed, one can proceed to identify the symmetries that an extended object breaks to derive its effective action. For instance, a point particle breaks spatial translations and boosts, while a spinning point particle breaks spatial translations and the full Lorentz group \cite{Delacretaz:2014oxa}. 

Due to the Goldstone's theorem \cite{Goldstone:1961eq} and the inverse Higgs constraint \cite{Ivanov:1975zq}, the symmetry breaking pattern for a spinning object implies the existence of a Nambu-Goldstone field, that takes the role of the angular velocity \cite{Delacretaz:2014oxa}, yielding a very natural construction for the action of spinning objects. Although it has been pointed out that this is a theory for "slowly" spinning compact objects, given that from construction the effective action corresponds to the low energy dynamics of the theory, we show that we can safely consider the current observed compact objects through GWs as slowly spinning.

Therefore, in this work we review and build on the EFT for spinning extended objects introduced in \cite{Delacretaz:2014oxa}. We derive in detail the building blocks of the effective theory and its constraints, which allows us to build up the tower of invariant operators and form the effective action to all orders.  We have extended the model by considering the electromagnetic charge U(1) symmetry, as an internal symmetry in the coset parameterization, which allows us to derive the Einstein-Maxwell action. Then, by identifying the symmetry breaking pattern of charged spinning object, we construct the corresponding corrections to the point particle. We show that the spin corrections considered in the effective theory for spinning extended objects in \cite{Levi:2015msa}, which are used to obtain state of the art results needed for GW extraction, are encoded in the allowed boson couplings.

For the internal structure of the objects, we take into account size \cite{Goldberger:2004jt}, and dissipative effects \cite{Goldberger:2005cd, Goldberger:2020fot}. The coefficients accompanying the operators of these effects, encode the microphysics of the object, and based on results in the literature, we can identify them without having to do the explicit computations. Tidal effects for the case of spinning compact objects have been taken into account in \cite{Porto:2005ac,Levi:2014gsa}, as well as for noncompact objects in \cite{Endlich:2015mke}, by considering only rotationally invariant operators and associating departures from sphericity with higher order corrections. The coefficients due to tidal effects  for compact objects can be matched from \cite{Poisson:2004cw, Chia:2020yla,Hinderer:2007mb, Yagi:2016bkt, Yagi:2016ejg}. 
We consider quasi-static dynamical tides on the NS as well, and match the coefficient to \cite{Chakrabarti:2013xza}.

We consider the size effects induced by charge as well, known as the polarizability, for which some of the coefficients are unknown. On the dissipative effects for spinning BHs, we consider the absorption of gravitational and electromagnetic waves \cite{Goldberger:2005cd}, as well as the dissipation generated by the spin \cite{Goldberger:2020fot,Porto:2007qi,Endlich:2015mke,Chia:2020yla}, for which coefficients have been matched by analytical means \cite{Goldberger:2005cd,Chia:2020yla}. For a NS, dissipation accounts for the energy loss due to the viscosity of the star, and state of the art hydrodynamical simulations are needed to match the coefficients.

In section \ref{sec:EFT},  we start with a brief introduction to effective field theory and the theoretical tools and concepts needed to develop this work. In section \ref{sec:electro}, we start with a pedagogical introduction deriving a classical theory of electromagnetism, as well as the effective action for charged spheres in flat space-time. This is done with particular emphasis to introduce the tools of the coset construction with a concrete and simple known example. In section \ref{sec:gravity}, we derive the Einstein-Maxwell action in the vierbein formalism, and then derive the covariant building blocks and constraints to build up the effective action of a charged spinning compact object in curved space-time, making the connection to the currently used theories that uses conjugate variables. Then, we match the coefficients of the effective theory for compact objects from the literature and comment about the properties of compact objects. Finally in section \ref{sec:discussion}, we conclude. 

\section{Effective Field Theory}
\label{sec:EFT}
In the observable universe there exists a wide variety of scales, from the quantum world to the cosmological structure. In this scale range we can find all kind of physical phenomena, for which the dynamics will be different from one scale to another. There is an important property of nature, known as decoupling, that has allowed us to understand the physical phenomena emerging at the different scales. The decoupling of physical effects means that most of the details, or short distance physics, are irrelevant for the description of larger distance phenomena.

The theoretical framework we have developed for modeling the laws of nature is known as field theory, both at quantum and classical level. Field theories also have the property that short distances mostly decouple from larger ones, and the set of tools that have been developed to show the symmetries, hierarchy of scales and decoupling of physical effects, are known as effective field theories. These tools, with the appropriate knowledge of the symmetry pattern, relevant scales and hierarchies, allows us to simplify the description of different systems.

In this section we introduce the concepts of EFTs by formulating an effective action for charged point particles in flat space-time. We derive the effective action using the coset construction.

\subsection{Conventions}
\label{subsec:conventions}

We differentiate between space-time and local Lorentz indices as in \cite{Delacretaz:2014oxa}:

\begin{itemize}
	\item $\mu, \nu, \sigma, \rho ...$ denote space-time indices.
	\item $a, b, c, d ...$ denote Lorentz indices.
	\item $i, j, k, l ...$ denote spatial components of the Lorentz indices.
\end{itemize}

We denote the time of occurrence and the location in space of an event with the four component vector, $x^a = (x^0,x^1,x^2,x^3) = (t,\vec{x})$, and define the flat space-time interval, $\mathrm{d}s$, between two events, $x^a$ and $x^a+\mathrm{d}x^a$, by the relation

\begin{equation}
\mathrm{d}s^2 = - c^2 \mathrm{d}t^2 + \mathrm{d} x^2 + \mathrm{d} y^2 + \mathrm{d} z^2 ,
\label{eq:space-timeinterval}
\end{equation}

\noindent which we write using the notation

\begin{equation}
\mathrm{d} s^2 = \eta_{ab}\mathrm{d}x^a\mathrm{d}x^b \,\mathrm{;} \,\,\,\,\,\;\;\;\;  \eta_{ab} = \mathrm{diag} (-1, +1, +1, + 1) .
\label{eq:lineinterv}
\end{equation}

\noindent In the last expression we have introduced the summation convention, which states that any index that is repeated in an expression, is summed over the range of values taken by the index. 

\subsection{Worldline Point Particle Dynamics}
\label{subsec:worldline}

The most important consideration when modelling compact objects with EFTs is that they can be treated as point particles with higher order corrections. To describe the dynamics of point particles, it is necessary to specify a continuous sequence of events in the space-time by giving the coordinates, $x^a (\sigma)$, of the events along a parametrized curve, defined in terms of a suitable parameter, $\sigma$. There is one curve among all possible curves in the space-time, which describes the trajectory of a material particle moving along some specified path, known as the worldline. We can consider it as a curve in the space-time, with $\sigma = c t$, acting as a parameter so that $x^a = (c t, \vec{x}(t))$. The existence of a maximum velocity $|\vec{u}| < c$, which requires the curve to be time-like everywhere, $\mathrm{d}s < 0 $, allows us to have a direct physical interpretation for the arc length along a curve. 

Consider a clock attached to the particle frame, or the proper frame $\tilde{\mathrm{F}}$, which is moving relative to some other inertial frame F on an arbitrary trajectory. As measured in the frame F, during a time interval between $t$ and $t + \mathrm{d}t$, the clock moves through a distance $|\mathrm{d}\vec{x}|$. The proper frame $\tilde{\mathrm{F}}$, which is moving with same velocity as the clock, will have $\mathrm{d}\vec{\tilde{x}} = 0$. If the clock indicates a lapse of time, $\mathrm{d}t \equiv \mathrm{d}\tau$, the invariance of the space-time interval, eq. (\ref{eq:space-timeinterval}), implies that 

\begin{equation}
\mathrm{d} s^2 = - c^2 \mathrm{d} t^2 + \mathrm{d} x^2 + \mathrm{d} y^2 + \mathrm{d} z^2 = \mathrm{d} \tilde{s}^2 = -c^2 \mathrm{d} \tau^2.
\end{equation}

\noindent Thus, we obtain the lapse of time in a moving clock

\begin{equation}
\mathrm{d}\tau = \mathrm{d}t\sqrt{1- \frac{v^2}{c^2}},
\end{equation}

\noindent along the trajectory of the clock. The total time that has elapsed in a moving clock between two events, known as the proper time $\tau$, is denoted as

\begin{equation}
\tau = \int \mathrm{d} \tau = \int^{t_2}_{t_1} \mathrm{d}t \sqrt{1 - \frac{v^2}{c^2}}.
\label{eq:tau}
\end{equation}

We now proceed to build the action of a free point particle. The Lagrangian must be constructed from the trajectory, $x^a(\tau)$, of the particle, and should be invariant under Lorentz transformations. The only possible term should be proportional to the integral of  $\mathrm{d} \tau$, yielding the action 

\begin{equation}
\mathcal{S} = -\alpha \int_{\tau} \mathrm{d} \tau = - \alpha \int  \sqrt{1 - \frac{v^2}{c^2}}  \mathrm{d}t ,
\end{equation}

\noindent where $\alpha$ is a dimensionful constant. To recover the action of a free point particle from non-relativistic mechanics, we take the limit $c \rightarrow \infty$, for which the Lagrangian yields $\mathcal{L} = \alpha v^2/2c^2$. By comparing our point particle Lagrangian with the non-relativistic one,  $\mathcal{L} = (1/2) mv^2$, with $m$ the mass of the particle, we find that, $\alpha=mc^2$. Thus, the action for a relativistic point particle is 

\begin{equation}
\mathcal{S} = - mc^2 \int \mathrm{d}\tau,
\end{equation}

\noindent which corresponds to the arc length of two connecting points in the space-time.

\subsection{Symmetries in Classical Field Theory}
\label{subsec:symmetries}

We consider symmetries that can be labelled by a continuous parameter, $\theta$. Working with the Lie algebras of a group $G$, we write a group element as a matrix exponential 

\begin{equation}
U = e^{i \theta^u T_u},
\end{equation}

\noindent where the generators $T_u$, with $u = 1,...,n$, form a basis of the Lie algebra of G. The $T$'s generators are hermitian if $U$ is unitary. For each group generator, a corresponding gauge field arises,  which in this case is the field $\theta$.

The properties of a group G are encoded in its group multiplication law

\begin{equation}
[T_u, T_v] = T_u T_v - T_v T_u  = i c_{uvw} T_w,
\label{eq:algebraT}
\end{equation}

\noindent where, $c_{uvw}$, are the structure constant coefficients. The last expression defines the Lie algebra of the group G. The Lie bracket is a measure of the non-commutativity between two generators.

It is also possible to consider in the local framework of field theory, continuous symmetries that have a position dependent symmetry parameters, $\theta = \theta (x)$. The space-time dependent symmetry transformation rules are called local or gauge symmetries. For global symmetries, $\theta$ do not depend on space-time position. There is also a distinction between internal symmetries and space-time symmetries, on whether they act or not on space-time position. An example of an internal symmetry, where $x$ is unchanged, is

\begin{equation}
\phi^u (x) \rightarrow U \phi^u (x) U^{-1} = \mathcal{U}_v^{\; u} \phi^v (x),
\end{equation}

\noindent while an example for a space-time symmetry is the transformation 

\begin{equation}
\phi^{u} (x) \rightarrow V \phi^{u} (x) V^{-1} = \mathcal{V}_{v}^{\; u} \phi^{v} (x'),
\end{equation}

\noindent with $x^{' u} = \mathcal{V}_v^{\; u} x^{v}$. Both internal and space-time symmetries can arise in global or gauged varieties. 

\subsubsection{Symmetries of Special Relativity}

The symmetry of Special Relativity is determined by the Poincaré symmetry. Its Lie group, known as the Poincaré group, G = ISO(3,1), is the group of Minkowski space-time isometries that includes all translations and Lorentz transformations.

\subsubsection*{The Lorentz Group}

The Lorentz group, SO(3,1), is the group of linear coordinate transformations

\begin{equation}
x^{a} \rightarrow x'^{a} = \Lambda^{a}_{\;\;b} x^{b},
\label{eq:coordtransf}
\end{equation}

\noindent that leave invariant the quantity

\begin{equation}
\eta_{a b} x^{a} x^{b} = -(ct)^2 + x_1^2 + x_2^2 + x_3^2,
\label{eq:dsi1}
\end{equation}

\noindent with $\mathrm{det}\Lambda =  1$. In order for eq. (\ref{eq:dsi1}) to be invariant, $\Lambda$ must satisfy

\begin{equation}
\eta_{a b} x'^{a} x'^{b} = \eta_{a b} (\Lambda^{a}_{\;\; c}x^{c})(\Lambda^{b}_{\;\; d}x^{d}) = \eta_{c d} x^{c} x^{d},
\end{equation}

\noindent which implies the transformation of the metric as

\begin{equation}
\eta_{c d}  = \eta_{a b} \Lambda^{a}_{\;\; c}\Lambda^{b}_{\;\; d}.
\label{eq:213}
\end{equation}

Consider an infinitesimal Lorentz transformation, with the Lorentz generators $J_{ab}$, and its corresponding field $\alpha_{ab}$. We can expand

\begin{equation}
\Lambda^{a}_{\;\; b} = (e^{\frac{i}{2} \alpha^{cd} J_{cd} })^a_{\;\; b} = (e^{\alpha})^a_{\;\; b} \approx \delta^{a}_{\;\; b} + \alpha^{a}_{\;\; b},
\label{eq:lambda}
\end{equation}

\noindent where the factor of, $\frac{1}{2}$, is taking into account the sum over all $a$ and $b$. From equation (\ref{eq:213}) we find

\begin{equation}
\alpha_{a b} = - \alpha_{ba},
\end{equation}

\noindent which is an antisymmetric 4x4 matrix with six components that are independent. Thus, the six independent parameters of the Lorentz group from the antisymmetric matrix, $\alpha_{ab}$, corresponds to six generators which are also antisymmetric $J^{ab} = - J^{ba}$.

Under Lorentz transformations, a scalar field is invariant,

\begin{equation}
\phi'(x') = \phi (x).
\end{equation}

\noindent A covariant vector field, $V^{a}$, transforms in a representation of the Lorentz group as

\begin{equation}
V^a \rightarrow (e^{\frac{i}{2} \alpha_{cd} J^{cd}})^{a}_{\;\;b} V^b,
\end{equation}

\noindent where the exponential is the matrix representation of the Lorentz group. If we consider an infinitesimal transformation, the variation of $ V^a$ reads, 

\begin{equation}
\delta V^{a} =  \frac{i}{2} \alpha_{c d} (J^{c d})^{a}_{\;\; b} V^{b},
\end{equation}

\noindent which is an irreducible representation. 

The explicit form of the matrix $(J^{ab})^{c}_{\;\; d}$, reads

\begin{flalign}
(J_{a b})^c_{\;\;d} = -i ( \delta^{c}_{\;\;a} \eta_{bd} - \delta^{c}_{\;\;b}  \eta_{ad} ).
\label{eq:J4in}
\end{flalign}

\noindent Using the form of the generator in eq. (\ref{eq:J4in}), we can compute the commutator

\begin{equation}
[J_{ab}, J_{c d}] = i( \eta_{a c} J_{b d} - \eta_{b c} J_{a d} + \eta_{b d} J_{a c} - \eta_{a d} J_{b c}),
\end{equation}

\noindent to find the Lie algebra, SO(3,1). The components of $J^{a b}$ can be rearranged into two spatial vectors 

\begin{equation}
J_{i} = \frac{1}{2} \epsilon_{ijk} J^{jk}, \;\;\; K^i = J^{i0},
\end{equation}

\noindent with, $J^{ij}$  and $K^i$, the generators of rotations and boosts, respectively.

The Lorentz group has six parameters: Three rotations in three $2D$ planes that can be formed with the $(x,y,z)$ coordinates that leave $ct$ invariant, which is the SO(3) rotation group, and three boost transformations in the $(ct,x)$, $(ct,y)$ and $(ct,z)$ planes that leave invariant $-(ct)^2+x^2$, $-(ct)^2+y^2$ and $-(ct)^2+z^2$, respectively. We parametrize the Lorentz matrix as

\begin{equation}
\Lambda^{0}_{\; \;0} = \gamma , \; \; \Lambda^{0}_{\; \; i} = \gamma \beta_i , \; \;  \Lambda^{i}_{\; \; 0} = \gamma \beta^i, \; \; \Lambda^{i}_{\; \; j} = \delta^{i}_{\;\; j} + (\gamma -1) \frac{\beta^i \beta_j}{\beta^2} , \; \;
\end{equation} 

\noindent with $\gamma = (1 - v^2/c^2)^{-1/2}$, the Lorentz factor, and $\beta^i$, the velocity 

\begin{equation}
\beta^i \equiv \frac{\eta^i}{\eta} \tanh \eta,
\label{eq:veltanh}
\end{equation}

\noindent where $\eta$ is the rapidity, defined as the hyperbolic angle that differentiates two inertial frames of reference that are moving relative to each other.

Therefore, the four vectors, $V^a$ and $V_a$, transforms under the Lorentz group as 

\begin{flalign}
V^{a} (x) \rightarrow V^{'a} (x') = \Lambda^{a}_{\;\; b} V^{b}(x), \;\;\;\;\;\; V_{a}(x) \rightarrow V_{a}^{'} (x') = \Lambda_{a}^{\;\; b} V_{b}(x),
\end{flalign} 

\noindent with $\Lambda_a^{\;\; b} = \eta_{ac} \eta_{bd} \Lambda^{c}_{\;\; d}$. The vectors are related via $V_a = \eta_{ab} V^{b}$. A tensor, $T^{ab}$, transforms as 

\begin{equation}
T^{a b} (x) \rightarrow T'^{a b} (x') = \Lambda^{a}_{\;\;c} \Lambda^{b}_{\;\;d} T^{c d} (x).
\end{equation}

\noindent In general, any tensor with arbitrary upper and lower indices transforms with a $\Lambda^{a}_{\;\; b} $ matrix for each upper index, and with $\Lambda_{a}^{\; \; b}$ for each lower one. We will simple denote Lorentz transformations as, $V^{'a} = \Lambda^{a}_{\;\;b} V^{b}$ and $T^{'ab} = \Lambda^{a}_{\;\;c} \Lambda^{b}_{\;\;d} T^{cd}$. 

\subsubsection*{The Poincaré Group}

To complete the Poincaré group, in addition to Lorentz invariance, we also require invariance under space-time translations. We can write a general element of the group of translations in the following form, 

\begin{equation}
U = e^{i  z^a P_{a}},
\end{equation}

\noindent where $z^a$ are the components of the translation,

\begin{equation}
x^{a} \rightarrow x^{a} + z^{a}, 
\label{eq:trans}
\end{equation}

\noindent and $P^{a}$ its generators. Lorentz transformations plus translations form the Poincaré group, ISO(3,1). The Poincaré group algebra reads

\begin{flalign}
[P_a, P_b] &= 0 \\
[P_a, J_{bc}] &= i(\eta_{ac} P_b  - \eta_{ab} P_c) \\ 
[J_{ab}, J_{c d}] &= i( \eta_{a c} J_{b d} - \eta_{b c} J_{a d} + \eta_{b d} J_{a c} - \eta_{a d} J_{b c}). 
\end{flalign}

\subsubsection{Gauge Symmetry of Classical Electromagnetism}

The gauge symmetry of classical electromagnetism is invariance under the U(1) gauge transformation. This is an internal symmetry for which the charge generator, Q, correspond to a time invariant generator, with its corresponding gauge field, $A_{\mu}(x)$, which is the electromagnetic gauge field. The local gauge symmetry is parametrized by a parameter $\theta =  \theta (x)$, and the group element is 

\begin{flalign}
U (x) = e^{\theta (x)}.
\end{flalign}

\noindent The gauge field, $A_{\mu}$, transforms under the U(1) symmetry as

\begin{flalign}
A_{\mu} (x) \rightarrow A_{\mu} (x) + \partial_{\mu} \theta (x).
\end{flalign}

\noindent Therefore, under Lorentz and U(1) transformations, the gauge field transforms as

\begin{flalign}
A_{\mu} (x) \rightarrow \Lambda_{\mu}^{\;\; \nu} A_{\nu} (x) + \partial_{\mu} \theta (x,\Lambda).
\end{flalign}

The commutations relations of the charge generator, $Q$, with the generators of the Poincaré group, are constrained by the Coleman-Mandula theorem \cite{Coleman:1967ad}. This theorem constrains the kinds of continuous space-time symmetries that can be present in an interacting relativistic field theory, and states that the most general possible transformations are determined by

\begin{equation}
U = \exp \left\{i\left(z^{a} P_{a} + i \sigma^a \mathcal{O}_a + \frac{i}{2} \omega^{a b} J_{a b}\right) \right\},
\end{equation}

\noindent with $P_{a}$, the generators of translations, $J_{a b}$, of Lorentz transformations, and $\mathcal{O}_a$, the rest of the generators. The generators, $\mathcal{O}_a$, must be from internal symmetries, and although they can fail to commute with themselves, $[\mathcal{O}_a, \mathcal{O}_b] \neq  0$, they must always commute with the space-time symmetry generators $[P_a, \mathcal{O}_b] = 0$ and $[J_{a b}, \mathcal{O}_c] = 0$. Nevertheless, the charge operator for the U(1) symmetry of electromagnetism, commutes with itself, thus obtaining the commutation relations: $[P_a, Q] = 0$, $[J_{a b}, Q] = 0$ and $[Q,Q] = 0$. 

\subsubsection{Transformation Properties of Gauge Fields}

The transformation properties of the gauge fields $\breve{e}_{\mu}^a$, $\breve{A}_{\mu}$ and $\breve{\omega}_{\mu}^{ab}$, under local translations, $e^{iz^a P_a}$, local Lorentz transformations, $e^{\frac{i}{2} \alpha_{cd} J^{cd}}$  and the local U(1) transformation, $e^{i \theta}$, read

\begin{flalign}
\begin{split}
U &= \;\,   e^{i \theta} \;:
\begin{cases}
\; \breve{A}_{\mu} \; \rightarrow \; \breve{A}_{\mu} - \partial_{\mu} \theta, \\
\; \breve{e}_{\mu}^{\; a} \;  \rightarrow \; \breve{e}_{\mu}^{\; a}, \\
\; \breve{\omega}_{\mu}^{ab} \rightarrow \; \breve{\omega}_{\mu}^{ab}.
\end{cases} \\
U &= \, e^{icP}:
\begin{cases}
\, \breve{A}_{\mu} \; \rightarrow \; \breve{A}_{\mu}, \\
\; \breve{e}_{\mu}^{\; a} \; \rightarrow \; \breve{e}^{\;a}_{\mu} - \breve{\omega}^a_{\mu b} z^b - \partial_{\mu} z^a, \\
\; \breve{\omega}_{\mu}^{ab} \rightarrow \; \breve{\omega}_{\mu}^{\;ab}.
\end{cases} \\
U &= e^{i \alpha J}:
\begin{cases}
\; \breve{A}_{\mu} \; \rightarrow \; \Lambda^{\;\; \nu}_{\mu} \breve{A}_{\nu}, \\
\; \breve{e}_{\mu}^{\; a} \; \rightarrow \;\, \Lambda^a_{\;b} \breve{e}^b_{\mu} = \; \breve{e}^{a}_{\mu} + \alpha^{a}_{\;b} \breve{e}^b_{\mu}, \\
\,\breve{\omega}_{\mu}^{ab} \,  \rightarrow \;\, \Lambda^{a}_{\; c} \Lambda^{b}_{\; d} \breve{\omega}^{cd}_{\mu} + \Lambda^a_{\;c} \partial_{\mu} (\Lambda^{-1})^{cb} = \breve{\omega}^{ab}_{\mu} + \breve{\omega}_{\mu}^{ac} \alpha^{b}_{\;c} + \breve{\omega}^{cb}_{\mu} \alpha^{a}_{\;c} - \partial_{\mu} \alpha^{ab}.
\end{cases} \\
\end{split},
\end{flalign}

\noindent where the indices are lowered and raised using the metric $\eta_{ab}$. The gauge field, $\breve{e}$, transforms inhomogeneously under local translations. Under Lorentz transformations,  $\breve{e}$ and $\breve{A}$, transforms linearly, while $\breve{\omega}^{ab}_{\mu}$ transforms as a connection. Under the U(1) transformation, only the gauge field, $\breve{A}$, transforms.

Finally, the electromagnetic field, the vierbein and spin connection, under diffeomorphisms transforms as 

\begin{flalign}
\begin{split}
A_{\mu}(x) \xrightarrow{\text{diffeo}}& A_{\mu}(x) - A_{\nu} (x) \partial_{\mu} \xi^{\nu} - \xi^{\nu} (x) \partial_{\nu} A_{\mu} (x),\\
e_{\mu}^a(x) \xrightarrow{\text{diffeo}}& \; e_{\mu}^a(x) - e_{\nu}^a (x) \partial_{\mu} \xi - \xi^{\nu} (x) \partial_{\nu} e_{\mu}^a (x),\\
\omega_{\mu}^{ab}(x) \xrightarrow{\text{diffeo}}& \; \omega_{\mu}^{ab}(x) - \omega_{\nu}^{ab} (x) \partial_{\mu} \xi - \xi^{\nu} (x) \partial_{\nu} \omega_{\mu}^{ab}.
\end{split}
\end{flalign}

\subsubsection{Noether's Theorem}

In Classical Field Theory, the symmetries of a Lagrangian and conserved quantities are closely related by Noether's theorem. This theorem allows us to obtain conserved quantities from the symmetries of the laws of nature. For instance, time translation symmetry gives conservation of energy, space translation symmetry gives conservation of momentum, rotation symmetry gives conservation of angular momentum, and the U(1) symmetry gives conservation of the electric charge .

Noether's theorem states that, on a classical solution of the equations of motion, there is a conserved current $j^{a}_{u}$, for each generator of a symmetry transformation \cite{Maggiore:2005qv}. Thus, for a field theory with the fields $\phi^a$, the $N$ currents $j^{a}_u$ are conserved

\begin{flalign}
	\partial_{a} j^{a}_u  = 0,
	\label{eq:noe}
\end{flalign}

\noindent where $u = 1,...,N$. We can define the charges $Q_u$

\begin{flalign}
	Q_u = \int \mathrm{d}^3 x j^{0}_u (x^a),
	\label{eq:conservedQ}
\end{flalign}

\noindent The current conservation implies that $Q_u$ is conserved.

\subsubsection{Spontaneous Symmetry Breaking}

In field theories, the ground state can fail to be invariant. In fact, any system other than vacuum will break at least some of the symmetries. If the ground state of a system is not invariant under a symmetry of its action, the symmetry is said to be spontaneously broken. If the symmetry is spontaneously broken, another state is produced once a transformation is applied to the ground state. The new state must have the same energy and is a candidate ground state. When the ground state is changed by a symmetry,  the spontaneously broken symmetries come with consequences that are described by Goldstone's theorem.

The Goldstone's theorem \cite{Goldstone:1961eq} summarizes very general implications for the low energy theory whenever the ground state of a system does not respect one of the global continuous symmetries of the system. It states that a system, for which a continuous global symmetry is spontaneously broken, the system must contain a state, $\ket{G}$, denoted as Nambu-Goldstone (NG) boson or Goldstone mode. This mode is created from the ground state, $\ket{\hat{0}}$, by a space-time independent symmetry transformation. $\ket{G}$ is defined by the condition that the matrix element

\begin{equation}
	\braket{G|j^0|\hat{0}} \neq 0,
	\label{eq:goldstoneboson}
\end{equation}

\noindent does not vanish, with $
j^0$ the density for the symmetry's conserved charge in eq. (\ref{eq:conservedQ}).  All properties of the Goldstone boson follow from the matrix element definition (\ref{eq:goldstoneboson}) and eq. (\ref{eq:conservedQ}). The Goldstone state is a symmetry transformation of the ground state, which makes it completely indistinguishable from the vacuum, thus creating a new ground state.

The Goldstone boson should decouple completely from all its interaction in the limit in which its momentum vanishes, such that the state $\ket{G}$ must be gapless. This means that its energy vanishes in the limit where the momentum vanishes,

\begin{equation}
	\lim_{p \rightarrow 0} E(p) = 0.
\end{equation} 

\noindent For a relativistic system, where $E(p) = \sqrt{p^2 + m^2}$, with $m$ the rest mass of the particle, the gapless condition implies the masslessness of the Goldstone particle. 

In the group theoretical approach, for the description of a symmetry breaking pattern where the group $G$ breaks to $H$, we can choose a basis for the generators such that we include the generators of $H$ as a subset, $\left\{T_u \right\} = \left\{t_v, X_{\alpha}\right\}$,
where the $t$'s generate the Lie algebra of $H$, and the $X_{\alpha}$ constitutes the rest.\footnote{We have chosen greek indices to differentiate from those of the unbroken group} The broken generators $X_{\alpha}$ typically do not also generate a group, but it can be regarded as generating the space of cosets $G/H$. We expect a Goldstone mode for each choice of $\alpha$.

The closure of $H$ under multiplication ensures 

\begin{equation}
	[t_u, t_v] = i c_{uvw} t_w,
\end{equation}

\noindent with no $X$'s on the right hand side. It is also possible to choose a basis of generators such that

\begin{equation}
	[t_u, X_{\alpha}] = i c_{u \alpha \beta} X_{\beta},
\end{equation}

\noindent with no $t$'s on the right-hand side. The implication of this, is that the $X_{\alpha}$'s falls into a representation of $H$, which when exponentiated to a finite transformation, can be expressed as linear transformation

\begin{equation}
	h X_{\alpha} h^{-1} = M_{\alpha}^{\;\;\beta}X_{\beta},
\end{equation}

\noindent for some coefficients $M_{\alpha}^{\;\; \beta}$, and for any $h = \exp(i \theta^u t_u) \in H$.

\subsection{The Coset Construction}
\label{sec:coset}

We start with the very basics of the coset construction to develop this paper. A brief but more comprehensive review can be found in \cite{Delacretaz:2014oxa,Penco:2020kvy}. We use the notation as in \cite{Delacretaz:2014oxa} to consider the breaking of internal \cite{Callan:1969sn, Coleman:1969} and space-time symmetries \cite{Volkov:1973vd, Ivanov:1981wn} alike.

The coset construction is a very general technique from the EFT framework that can be used whenever there is a symmetry breaking. The breaking of some of the symmetries implies the existence of additional degrees of freedom, known as Nambu-Goldstone bosons or simply as Goldstone fields.\footnote{Goldstone theorem \cite{Goldstone:1961eq} implies the existence of a Goldstone field for each broken internal symmetry, but for the case in which space-time symmetries are broken, there can be a mismatch on the number of degrees of freedom and broken symmetries, for which additional constraints are needed. See the Inverse Higgs constraint below.} The coset construction is then used to derive building blocks for the Goldstone fields that transform correctly under the relevant symmetries, blocks that can be used to build up invariant operators to form an effective action. Any state other than vacuum breaks at least some of the symmetries, and by appropriately identifying the pattern of the symmetry breaking, we can use it as a guide to derive the effective action.

We can formulate an EFT using the symmetry breaking pattern as the only input, knowing the full symmetry group G that is broken, and the
subgroup H that is non-linearly realized \cite{Penco:2020kvy}. If the group is broken, G $\rightarrow$ H, due to a spontaneous symmetry breaking, the coset recipe \cite{Delacretaz:2014oxa,Penco:2020kvy} tells us that we can classify the generators into three categories:

\begin{flalign}
\begin{split}
P_{a} &= \mathrm{generators \; of \; unbroken \; translations},\\
T_A &= \mathrm{generators \; of \; all \; other \; unbroken \; symmetries}, \\
X_{\alpha} &= \mathrm{generators \; of \; broken \; symmetries},
\end{split}    
\end{flalign}

\noindent where the broken generators, $X_{\alpha}$, and the unbroken ones, $T_A$, can be of space-time symmetries, as well as of internal ones.  Whenever the set of generators for broken symmetries is non-zero, some Goldstone fields will arise. Thus, we must build an effective action for the Goldstone fields that is invariant under the whole symmetry group of consideration. The power of the coset construction, is that we can formulate an invariant EFT in which the broken symmetries and the unbroken translations are realized non-linearly on the Goldstone fields \cite{Delacretaz:2014oxa}. 

Following the coset recipe \cite{Delacretaz:2014oxa,Penco:2020kvy}, we do a local parametrization of the coset, G/H$_0$, with $H_0$, the subgroup of $H$ generated by the unbroken generators, $T$'s. The coset is parametrized as 

\begin{flalign}
g (x, \pi) = e^{iy^{a}(x) P_{a}} e^{i \pi^{\alpha}(x) X_{\alpha}},
\label{eq:gcoset}
\end{flalign}

\noindent where the factor, $e^{iy^a (x) P_a}$, describes a translation from the origin of the coordinate system to the point, $x_a$, at which the Goldstone fields, $\pi^{\alpha} ( x )$, are evaluated. This factor ensures that the $\pi$’s transform correctly under spatial translation. The group element $g$, which is generated by the $X$'s and the $P$'s, is known as the coset parametrization. For the case of flat space-time, the translation is simply parametrized by, $e^{ix^a P_a}$, with $y(x) \equiv x$.

To derive the building blocks that depend on the Goldstone bosons and that have simple transformation rules, we first note that the Goldstone fields, when appearing in the Lagrangian, they
are coupled through its derivatives. Then, we introduce the Maurer-Cartan form, $g^{-1} \partial_{\mu} g$, a very convenient quantity that is an element of the algebra
of G, and that can be written as a linear combinations of all the generators \cite{Delacretaz:2014oxa, Penco:2020kvy},

\begin{equation}
g^{-1} \partial_{\mu} g = ( e_{\mu}^{\; \; a} P_a + \nabla_{\mu} \pi^{\alpha} X_{\alpha} + C_{\mu}^{\; \; B} T_B).
\label{eq:mauren}
\end{equation}

\noindent The coefficients $e_{\mu}^{a}$, $\nabla_{\mu} \pi^{\alpha}$ and $C_{\mu}^B$, in general are non-linear functions of the Goldstones, and are basic ingredients of the effective theory, with $\nabla_{\mu} \pi^{ \alpha} = e_{\mu}^{\; \;a} \nabla_{a} \pi^{\alpha}$ and $C_{\mu}^{\; \; B} = e_{\mu}^{\; \; a} C_{a}^{\;\; B}$. The explicit expression of the aforementioned ingredients can be obtained using the algebra of the group $G$. 

Following the coset recipe \cite{Delacretaz:2014oxa, Penco:2020kvy}, we can use the coefficients of the unbroken symmetries, $C$'s, and its operators, $T$'s, to define the covariant derivative,

\begin{flalign}
\nabla_a \equiv (e^{-1})_{a}^{\; \mu} (\partial_{\mu} + i C_{\mu}^B T_B).
\end{flalign}

\noindent This covariant derivative can be used to define higher covariant derivatives on the Goldstone fields, as well as on some of the building blocks and additional fields that transform linearly under the unbroken group. Then, by considering all the allowed contractions of the building blocks and their higher covariant derivatives, it is possible to build up an invariant effective action under the full symmetry group $G$.

In gauge symmetries, it is necessary to promote the partial derivative to a covariant one in the Maurer-Cartan form, $\partial_{\mu} \rightarrow D_{\mu}$ \cite{Delacretaz:2014oxa}. Consider the gauged generator, $E_I$, from a subgroup, $G' \subseteq G$, with corresponding gauge field, $w^{I}_{\mu}$. Thus, by replacing the partial derivative with a covariant one, we obtain the modified Maurer-Cartan form, 

\begin{equation}
g^{-1} \partial_{\mu} g \rightarrow g^{-1} D_{\mu} g = g^{-1} (\partial_{\mu} + i w^{I}_{\mu} E_{I} )g
\label{eq:maurenmod}.
\end{equation}

\noindent This modification of the Maurer-Cartan form can also be written as a linear combination of the generators as in eq. (\ref{eq:mauren}), with a new building block made up of the gauge field, $w_{\mu}^{I}$, accompanying the gauged generator, $E_I$. Now the building blocks can also depend on the included gauge fields. The modified Maurer-Cartan form, $ g^{-1} D_{\mu} g $, is invariant under local transformations, and its explicit components can be obtained using the commutation relations of the generators.

\subsection*{Inverse Higgs Constraint}

The Goldstone’s theorem \cite{Goldstone:1961eq}, which states that a Goldstone mode exists for each broken generator, is only valid for internal symmetries. If space-time symmetries are spontaneously broken, there can be a mismatch in the number of broken generators and the number of bosons \cite{Low:2001bw}. Nevertheless, we can preserve all the symmetries by imposing additional local constraints, which can be solved to write down some of the Goldstone’s modes in terms of others \cite{Penco:2020kvy}. Using the inverse Higgs constraint \cite{Ivanov:1975zq}, we can set to zero one or more of the coset covariant derivatives,  whenever $X$ and $X'$, are two multiplets of the broken generator, such that the commutators of the unbroken translations, $P$, and the broken generator, $X'$, yields a different broken generator, $X$: $\; [P, X'] \supset X$. If this is the case, we can set some of the covariant derivatives of the Goldstones to zero. By imposing all possible inverse Higgs constraints, one obtains the only relevant building blocks.

\section{Electrodynamics of Spheres}
\label{sec:electro}

\subsection{Classical Electromagnetism} 

It is well known that a classical theory of electromagnetism obeys the symmetries of special relativity determined by the Poincaré group, ISO(3,1), as well as the symmetries of the U(1) charge symmetry group. Thus, the full group is, G = U(1)$\times$ISO(3,1), which contains the generators for Lorentz transformations, $J^{ab}$, and gauge field, $\breve{\omega}_{\mu}^{ab}$, the generators of translations, $P_a$, and gauge field, $\breve{e}^{a}_{\mu}$, and the generator of charge, $Q$, and a gauge field $\breve{A}^{}_{\mu}$.\footnote{Note that we have defined the gauge fields, $\breve{o}$'s, compared to \cite{Delacretaz:2014oxa}, in which the gauge fields are defined as $\tilde{o}$'s. We will reserve the tilde to refer to quantities in the comoving frame.} Charge corresponds to a time invariant generator of the internal symmetry group, U(1), while the symmetry of the Poincaré group is a space-time symmetry (See sec \ref{subsec:symmetries} for more details).

Therefore, we parametrize the coset, U(1)$\times$ISO(3,1)/U(1)$\times$SO(3,1), separating translations from the rest of the group. The coset parametrization in flat space-time is given by the group element 

\begin{flalign}
	g = e^{i x^a \bar{P}_a} = e^{ix^a P_a} e^{i x^{0} Q} ,
\end{flalign}

\noindent which contains all the unbroken translation, $\bar{P}_a = \bar{P}_0 + P_i = P_a + Q $, with $\bar{P}_0 = P_0 + Q$. Then, the Maurer-Cartan form reads

\begin{flalign}
\begin{split}
g^{-1} D_{\mu} g &=  e^{-i x^a \bar{P}_a} \left( \partial_{\mu} + i \breve{A}_{\mu}^{} Q + i \breve{e}_{\mu}^{\;\; a} P_a + \frac{i}{2} \breve{\omega}_{\mu}^{ab} J_{ab} \right) e^{ix^a \bar{P}_a}  \\
&= \partial_{\mu} +  i A_{\mu} Q + i e_{\mu}^{\;\;a} P_a + \frac{i}{2} \omega_{\mu}^{ab} J_{ab},
\label{eq:maurercartanelectro}
\end{split}
\end{flalign}

\noindent where we have used the commutation relation rules of the symmetries (See appendix \ref{subsec:symmetries}) to obtain  

\begin{flalign}
e_{\mu}^{\;\; a} &= \breve{e}_{\mu}^{\;\;a} + \partial_{\mu} x^a + \breve{\omega}_{\mu}^{ab} x_b, \label{eq:efieldflat}\\
A_{\mu} &=   \breve{A}_{\mu} + \partial_{\mu} \xi (x), \\
\omega^{ab}_{\mu} &= \breve{\omega}^{ab}_{\mu}.
\end{flalign} 

Before going any further, we consider the case of flat space-time. From a geometrical perspective, this limit implies that the curvature and the torsion tensor are equal to zero, which can be measured from the commutator of the two covariant derivatives, $[D_{\mu}, D_{\nu}]$. By considering the covariant derivative appearing in the Maurer-Cartan form, (\ref{eq:maurercartanelectro}), we can cast the commutator as, 

\begin{flalign}
\begin{split}
[D_{\mu}, D_{\nu}] =&  i \breve{F}_{\mu \nu}^{} Q +  i \breve{T}^{a}_{\mu\nu} P_a + \frac{i}{2} \breve{R}^{ab}_{\mu \nu } J_{ab}, \\
=& i\left(\partial_{\mu} \breve{A}^{}_{\nu} - \partial_{\nu} \breve{A}^{}_{\mu} \right)Q + i \left(\partial_{\mu} \breve{e}^{a}_{\;\;\nu} - \partial_{\nu} \breve{e}^{a}_{\;\;\mu}  + \breve{e}^{}_{\mu b} \breve{\omega}^{ab}_{\nu} - \breve{e}^{}_{\nu b} \breve{\omega}^{ab}_{\mu}\right)P_a\\
&+ \frac{i}{2} \left( \partial_{\mu} \breve{\omega}^{ab}_{\nu} -\partial_{\nu} \breve{\omega}_{\mu}^{ab} + \breve{\omega}^{a}_{\mu c} \breve{\omega}^{cb}_{\nu} - \breve{\omega}^{a}_{\nu c} \breve{\omega}^{cb}_{\mu} \right) J_{ab}.
\label{eq:commutatorflat}
\end{split}
\end{flalign}

\noindent Given that in flat space-time, $\breve{T}_{a}^{\mu \nu} = 0 $ and $\breve{R}_{ab}^{\mu \nu} = 0$, it implies that 

\begin{flalign}
\breve{\omega}^{ab}_{\mu} = 0, \;\;\;\;
\end{flalign}

\noindent for which eq. (\ref{eq:efieldflat}), takes the simple form  

\begin{flalign}
e^{\;\; a}_{\mu} = \partial_{\mu} x^a = \delta_{\mu}^{\;\; a}.
\end{flalign}

The field, $e^{\; a}_{\mu}$, is known as the vierbein, which is used for deriving an invariant volume element $\mathrm{d}^4 x \, \mathrm{det}\,e$, and which for the case of flat space-time is simply $\mathrm{d}^4 x$.  Given the coset recipe, we can use the coefficients of the unbroken U(1) generator in eq. (\ref{eq:maurercartanelectro}), to define the gauge covariant derivative, 

\begin{flalign}
    \nabla_a^{} \equiv  \partial_a + i Q A_a ,
\end{flalign}

\noindent which is the usual covariant derivative in a classical field theory of electromagnetism. Having defined the vierbein and the gauge covariant derivative, we can proceed to build invariant actions \cite{Delacretaz:2014oxa}. 

The first building block can be extracted from the covariant commutator of the covariant derivatives, eq. (\ref{eq:commutatorflat}), which in the flat space-time limit,

\begin{flalign}
g^{-1}[D_{\mu},D_{\nu}]g = i(\partial_{\mu} A_{\nu} - \partial_{\nu} A_{\mu} )Q = i F_{\mu \nu} Q,
\label{eq:commg} 
\end{flalign}

\noindent with $F_{\mu \nu } = \partial_{\mu} A_{\nu} - \partial_{\nu} A_{\mu}$, the electromagnetic tensor. Thus, the first invariant term that can be build with our covariant building block, $F_{\mu \nu}$, is,

\begin{flalign}
\mathcal{S} = -\alpha \int \mathrm{d}^4 x F_{\mu \nu} F^{\mu \nu }.
\end{flalign}

\noindent The coefficients of the theory can be treated as free parameters that are to be fixed by experiments or from the literature. Thus, to reproduce Maxwell's theory, we find the coefficient, $\alpha = (4 \mu_0)^{-1}$, with $\mu_0$, the  magnetic permeability of vacuum, such that we have the well known Maxwell action,

\begin{flalign}
\mathcal{S} = - \frac{1}{4 \mu_0} \int \mathrm{d}^4 x F_{a b} F^{a b },
\label{eq:maxwell}
\end{flalign}

\noindent where we have used, $ F^{ab} =  e^{\; \; a}_{\mu} e^{\; \; b}_{\nu} F^{\mu \nu}$. Eq. (\ref{eq:maxwell}) lives in the bulk, while the action that describes the sphere, will live in the worldline.

\subsection{Charged Spheres}

With the underlying theory of classical electromagnetism, we identify the symmetry pattern for a charged point particle, and derive the building blocks to construct an effective action. Then, by considering spherical objects at rest, and departures from sphericity, we add size effects to describe charged spheres.  

\subsubsection*{Charged Point Particles}

In general, a point particle breaks spatial translations, and boosts by choosing a preferred reference frame. Under a U(1) symmetry, the state of a charged point particle is an eigenstate of the charge, and does not break the U(1) symmetry.  The symmetry breaking pattern reads

\begin{flalign}
\begin{split}
\mathrm{Unbroken \; generators} &=
\begin{cases}
&\bar{P}_0 \equiv P_0 + Q \;\;\;\;\;\;\;\;\;\, \mathrm{time \; translations},\\
& J_{ij} \;\;\;\;\;\;\;\;\;\; \;\;\;\;\;\;\;\;\;\; \;\;\;\;\; \mathrm{spatial \; rotations.}	
\end{cases}\\
\mathrm{Broken \; generators} &= 
\begin{cases}
&P_i \;\;\;\;\;\;\;\;\;\;\;\;\;\;\;\;\;\;\;\;\;\;\;\;\;\; \mathrm{spatial \; translations} , \\ 
&J_{0i} \equiv K_i \;\;\;\;\;\;\;\;\;\;\;\;\;\;\; \; \mathrm{boosts},
\end{cases}
\end{split}
\end{flalign}

\noindent where we have included the charge generator as a time invariant generator of translations \cite{Delacretaz:2014oxa}. Given this pattern, we can parametrize the coset as  

\begin{flalign}
g = e^{i x^a (\lambda) \bar{P}_a} e^{i \eta^i (
	\lambda) K_i} = e^{i x^a (\lambda) \bar{P}_a } \bar{g}.
\label{eq:cosetcharge}
\end{flalign}

\noindent with $\lambda$, the worldline parameter that traces out the trajectory of the particle, $\eta^i$, the Goldstone mode, and $\bar{g} = e^{i \eta^i (\lambda) K_i}$. 

The important building blocks are contained in the Maurer-Cartan form projected into the particle's trajectory, $\dot{x}^{\mu} g^{-1} D_{\mu} g$. It can be casted as a linear combination of the generators,

\begin{flalign}
\begin{split}
\dot{x}^{\mu} g^{-1} D_{\mu} g &= \dot{x}^{\mu} g^{-1} (\partial_{\mu} + i \breve{A}_{\mu} Q + i \breve{e}_{\mu}^{\;\; a} P_a ) g \\
&= \dot{x}^{\mu} \bar{g}^{-1} (\partial_{\mu} + i A_{\mu}^{} Q + i e_{\mu}^{\;\; a} P_a ) \bar{g}\\
&= i \dot{x}^{\mu} (  A_{\nu}^{} \Lambda_{
\;\;\mu}^{\nu} Q  +   e^{\;b}_{\mu} \Lambda_{b}^{\; 0} P_0 +  e^{\;b}_{\mu} \Lambda_{b}^{\; i} P_i) + i (\Lambda^{-1})^0_{\;c} \dot{\Lambda}^{ci} K_i,\\
&= i E \left( P_0 + A Q + \nabla \pi^i P_i +  \nabla \eta^i K_i \right), 
\label{eq:lbuildingflatcharge}
\end{split}
\end{flalign}

\noindent  where the dot means derivative with respect to the worldline parameter, $\lambda$, and $\Lambda^a_{\;\; b} = \Lambda^a_{\;\; b} (\eta) \equiv (e^{i \eta^j K_j})^{a}_{\;\;b}$, being the boost matrix of the Lorentz transformations, which is a function of the Goldstone field, $\eta^i$. Thus, the covariant quantities are

\begin{flalign}
\begin{split}
E &= \dot{x}^{\mu} e_{\mu}^{\; \; b} \Lambda_b^{\;\; 0} ,\\
A &=  E^{-1} \dot{x}^{\mu}  A_{\nu}^{}  \Lambda_{\;\; \mu}^{\nu},  \\
\nabla \pi^i &= E^{-1} \dot{x}^{\mu} e_{\nu}^{\; \; b} \Lambda_b^{\;\; i}, \\
\nabla \eta^i &= E^{-1} (\Lambda^{-1})^0_{\;c} \dot{\Lambda}^{ci}.
\label{eq:buildingbflat}
\end{split}
\end{flalign}

\noindent These are some of the building blocks that we will use to build up an invariant action. Nevertheless, as pointed out before, given that we are working with space-time symmetries, there are some subtleties with the counting of Goldstone fields, for which the Inverse Higgs constraint can be placed. 

In this case, the commutator between boosts and unbroken time translations gives broken spatial translations \cite{Delacretaz:2014oxa}. Therefore, by imposing the inverse Higgs constraint, we set to zero the covariant derivative, $\nabla \pi^i = 0$, such that we obtain the relation

\begin{flalign}
\nabla \pi^i = e^{-1} \dot{x}^{\mu} e_{\mu}^{\;\; b} \Lambda_b^{\;\; i} = e^{-1} (\dot{x}^{\mu} e_{\mu}^{\;\;0} \Lambda_{0}^{\;\; i} + \dot{x}^{\mu} e_{\mu}^{\;\;} \Lambda_{j}^{\;\; i}) = 0.
\label{eq:inverseHflat}
\end{flalign}

\noindent Solving for this constraint, we obtain the velocity (See eq. \ref{eq:veltanh}) 

\begin{flalign}
\beta^i \equiv \frac{\eta^i}{\eta} \tanh \eta  = \frac{\dot{x}^{\nu} e_{\nu}^{\;\; j}}{\dot{x}^{\nu} e_{\nu}^{\;\; 0}},
\end{flalign}

\noindent where we have used the fact that, in flat space-time, $e_{\nu}^{\; \; a} = \delta_{\nu}^{\;\; a}$. Using the inverse Higgs constraint we have expressed the $\eta$'s in terms of the $\pi$'s. This result can be interpreted as the Goldstones, $\eta$'s, or in terms of the, $\beta$'s, as parametrizing the boost necessary to get into the moving particle rest frame \cite{Delacretaz:2014oxa}. 

We can also rewrite the inverse Higgs constraint in a way that the physical interpretation is transparent \cite{Delacretaz:2014oxa}. This is done by expressing eq. (\ref{eq:inverseHflat}), in terms of the  velocity in the proper frame, $u^a$, as

\begin{flalign}
u^a \Lambda_a^{\;\; i}  = 0,
\end{flalign} 

\noindent and defining the set of Lorentz vectors, $\hat{n}^a_{\;\; (b)}$, 

\begin{equation}
{\hat{n}^a_{\;\; (0)} \equiv u^a  = \Lambda^a_{\;\; 0} (\eta) \;, \;\;\;\; \hat{n}^{a}_{(i)} \equiv \Lambda^a_{\;\; i} (\eta)   }.
\end{equation}

\noindent This set of Lorentz vectors define an orthonormal basis in the particle's comoving frame.

Then, we analyse the covariant derivative of the boost Goldstone, $\eta^i$, in eq. (\ref{eq:buildingbflat}), which can be rewritten as 

\begin{flalign}
\nabla \eta^i = \hat{n}^{(i)}_{a} \partial_{\tau} u^a = \hat{n}^{(i)}_{b} e_{\mu}^{\;b} a^{\mu} = a^i,
\end{flalign} 

\noindent with, $a^i $, the acceleration of the particle in the comoving frame. Thus, the covariant derivatives, $\nabla \eta^i$, corresponds to the component of the acceleration projected on the i-th vector defined by the set of Lorentz vectors, $\hat{n}^a_{(b)}$, measured in the proper frame of the particle \cite{Delacretaz:2014oxa}. In the absence of external forces, $\nabla \eta^i = 0$. Nevertheless, this building block is needed to take into account for a full description of the system, i.e. when the charge from an external body is relevant, such that the acceleration is nonzero.

Thus, we are left with the building blocks, 

\begin{flalign}
\begin{split}
e &= \dot{x}^{\mu} e_{\mu}^{\; \; 0}  ,\\
A &=  e^{-1} \dot{x}^{\mu}  A_{\mu}^{} ,  \\
\nabla \eta^i &= e^{-1}  a^i,
\label{eq:buildingfflat}
\end{split}
\end{flalign}

\noindent which are to be used in order to build an invariant action. We start with the building block, $e =|e| = \sqrt{e^2}$, and rewrite it as

\begin{flalign}
\begin{split}
|e| &= \sqrt{(e \nabla \pi^i)^2 + (e^2 - (e \nabla \pi^i)^2)}\\ 
&= \sqrt{(e \nabla \pi^i)^2 - (\dot{x}^{\nu} e_{\nu}^{\;\;a}  \dot{x}^{\mu} e^{\;\; b}_{\mu}  )},\\
&=\sqrt{-\eta_{ab} e_{\nu}^{\;\; a} e_{\mu}^{\;\; b} \dot{x}^{\nu} \dot{x}^{\mu}}  =\sqrt{-\eta_{ab}  \dot{x}^{a} \dot{x}^{b}} = \frac{\mathrm{d} \tau}{\mathrm{d} \sigma},
\end{split} 
\end{flalign}

\noindent where we have imposed the inverse Higgs constraint, and used the orthogonality property of the boost matrices, $\Lambda_{a}^{\;\; b} \Lambda^{a}_{\;\; c} = \delta^{b}_{\; \; c}$,  and $e^{\;\; a}_{\mu} = \delta^{\;\; a}_{\mu}$. We can build an invariant term made up of the building block $e$, 

\begin{equation}
\mathcal{S} = c_E \int \mathrm{d}\sigma e = c_E \int \mathrm{d}\sigma \frac{\mathrm{d}\tau}{\mathrm{d}\sigma}  = -m c^2 \int \mathrm{d} \tau,
\label{eq:ppflat}
\end{equation} 

\noindent and match the coefficient, $c_E = -mc^2$, to the action of a free relativistic point particle with mass, $m$. The last expression shows the action for a free relativistic point particle to all orders in the absence of external forces \cite{Delacretaz:2014oxa}.

With the covariant block, $A$, we add the point particle correction due to charge as,

\begin{equation}
\mathcal{S} = \int \mathrm{d} \sigma e ( -mc^2  + c_A A ) = \int \mathrm{d} \tau ( -mc^2  + qu^{a} A_{a}),
\label{eq:chargedpp}
\end{equation} 

\noindent with, $q$, the net charge of the charged particle, where the coefficient, $c_{A}$, has been matched from the classical theory of electrodynamics. Therefore, equation (\ref{eq:chargedpp}), is the action in the proper frame that describes charged point particles in classical electromagnetism. We now move into the description of size effects, to describe a charged sphere. 

\subsubsection*{Size Effects}

In the context of the EFT for extended objects, the addition of size  effects in the action of a point particle was introduced in \cite{Goldberger:2004jt}, which can be done in a systematic fashion. Size corrections must respect the symmetries of the theory, Lorentz and gauge U(1) symmetry, and thus contained in an infinite series expansion of all possible invariant operators made up of the covariant building blocks. To build up invariant terms made out of $F_{ab}$, we first define its transformation properties under local Lorentz transformations.  By considering the Lorentz parametrization of eq. (\ref{eq:cosetcharge}), $g_{L}$,  the electromagnetic tensor transforms as

\begin{equation}
F \equiv g^{-1}_{L}  \; \tilde{F},
\end{equation}  

\noindent such that $F$ is transformed in a linear representation under a Lorentz transformation as expected. The explicit transformation reads

\begin{equation}
F_{ab} = (\Lambda^{-1} )_{a}^{\;\; c}  (\Lambda^{-1} )_{b}^{\;\; d}  \tilde{F}^{}_{cd}.
\end{equation}

Having the correct transformation rules, we can form rotationally invariant objects made out of $F_{a b}$, $u^{a}$, and $a^{b}$, to build up the invariant action. These corrections in the action reads,

\begin{flalign}
	\mathcal{S} = \sum_n \int \mathrm{d} \tau  c_n \mathcal{O}_{n} (F_{a b}, u_{c}, a_{d}),
\end{flalign} 

\noindent with, $\mathcal{O}_n$, the invariant operators, $c_n$, its corresponding coefficients, and $n$, being chosen to the desired accuracy. The leading order, electric parity terms are

\begin{flalign}
	\mathcal{S} = \int \mathrm{d}\tau \left( n_qF^{b a} F_{c a} u_{b} u^{c}  + n_{qa} u_{a} a_{b} F^{a b} + .\;.\;. \right),
	\label{eq:size-electro}
\end{flalign}

\noindent where the ellipsis denotes higher order corrections. The first term correspond to the induced, electric parity, dipolar moment, while the second term can be seen as a higher order correction to the latter due to the acceleration of the body. 

The electric parity building block is identified, $E_{a} = F_{ba} u^b$ \cite{Goldberger:2005cd}, such that the first term in the last equation, $\propto E^a E_a$.  For the moment we restrict to operators with electric parity only. Nevertheless, a term with magnetic parity, $B_{a} = \frac{1}{2} \epsilon_{abcd} F^{bc} u^{d} $, can be build as well, $\propto B^{a} B_{a}$. The leading order corrections in (\ref{eq:size-electro}), with its corresponding coefficients, $c_{q} $, and $c_{qa} $, encode the short distance information, or the size structure, which are responsible for the polarization. The polarizability accounts for the deformation of the sphere in the presence of an external electromagnetic field. In general, the coefficients, $c_{q}$ and $c_{qa}$, are dependent on the radius of the sphere, therefore taking into account for the fact that a sphere is an extended object.  The specific form of these coefficient for a charged sphere is beyond the scope of this work.   

\subsubsection*{Dissipative Effects}

Dissipation in the EFT of extended objects was introduced in \cite{Goldberger:2005cd}. Dissipative effects of the sphere takes into account for its absorption of electromagnetic waves. These large number of degrees of freedom can be encoded in operators allowed by the symmetries of the object. For instance, for a rigid sphere, the allowed operators for rotations, SO(3), as well as parity eigenvalue, allows us to build the action \cite{Goldberger:2005cd} 

\begin{flalign}
    S =  \int \mathrm{d} \tau \mathcal{P}^a (\tau) u^b F_{ab},
\end{flalign}

\noindent with $\mathcal{P}(\tau)$, a composite dynamical operator corresponding to the electric parity, electromagnetic dipole moment. The specific form of the operator, $\mathcal{P}(\tau)$, and its coefficient are discussed in detail in the next section.

\subsection{The Effective Action}

Finally, by considering, the electric parity only, we write down the leading order effective action of a charged sphere in a classical theory of electromagnetism,

\begin{flalign}
	\mathcal{S}_{eff} = \int \mathrm{d} \tau\left(-m c^2 + q u^{a} A_{a}  + c_q  E^{a} E_{a } + c_{qa} E^a a_a + \mathcal{P}^a  F_{ab} u^b + .\,.\, . \, \right) + \mathcal{S}_0 ,
	\label{eq:effectactelectro}
\end{flalign}

\noindent with

\begin{flalign}
    \mathcal{S}_0 = - \int \mathrm{d}^4 x \frac{1}{4 \mu_0} F_{a b} F^{a b}.
    \label{eq:interactionelectro}
\end{flalign}

\noindent Eq. (\ref{eq:effectactelectro}) describes our effective theory. We have developed a theory of classical electromagnetism and described a charged sphere as a charged point particle with the finite size structure encoded in higher order operators in the action. In this approach, the description of the sphere lives in the worldline, while the interaction, eq. (\ref{eq:interactionelectro}), lives in the bulk.

\section{Compact Objects in General Relativity}
\label{sec:gravity}

Following the same methodology as in the previous section, before constructing a theory for compact objects, we review how a theory of gravity can be derived from the coset construction as in \cite{Delacretaz:2014oxa}, where a frame independent generalization of General Relativity, known as Einstein's vierbein field theory, is derived. The difference of our construction compared to the one in \cite{Delacretaz:2014oxa}, is the inclusion of the gauge symmetry group, $U(1)$, which allows us to derive the Einstein-Maxwell action in the vierbein formalism, as well as the correction to the point particle due to charge. Once the underlying theory of gravity has been developed, then we derive the leading order action for charged spinning compact objects in the effective theory of General Relativity (vierbein formalism). This is a theory that can naturally incorporate spinning objects.

\subsection{Effective Theory of Gravity}

There are two symmetries in gravity to consider: Diffeomorphisms invariance, and Poincaré symmetry, determined by the Poincaré group, ISO(3,1), which contains the generators for translations, $P_a$, and Lorentz transformations, $J_{ab}$, with their corresponding gauge fields, $\breve{e}_{\mu}^{a}$ and $\breve{\omega}_{\mu}^{ab}$. Both of the aforementioned symmetries of the system can be separated by considering the principal bundle, P(M,G), with base manifold, M, and structure group, G. In this way, we realize the matter fields as sections of their respective fiber bundle \cite{Delacretaz:2014oxa}. The coordinates, $x^{\mu}$, describing the position on the considered manifold, M, are not affected by the local Poincaré group, but it is transformed under diffeomorphisms. The local Poincaré transformations act along the fiber, while diffeomorphisms can be considered as relabeling the points on the manifold. 

To incorporate electrodynamics in the theory of gravity \cite{Delacretaz:2014oxa}, we add the U(1) symmetry of electromagnetism with its gauge field, $\breve{A}_{\mu}$, and generator $Q$, and proceed with the coset construction by gauging the Poincaré group and realizing translations non-linearly \cite{Ivanov:1981wn}. The coset then reads, U(1) $\times$ ISO(3,1)/U(1) $\times$ SO(3,1), with the coset parametrization

\begin{equation}
g = e^{i y^a (x) \bar{P}_a},
\label{eq:gravparam}
\end{equation}

\noindent where $\bar{P}_a = \bar{P}_0 + P_i = P_a + Q $. We now compute the Maurer-Cartan form from the coset parametrization (\ref{eq:gravparam}), which is expressed as a linear combination of the generators of the theory,

\begin{flalign}
\begin{split}
g^{-1} D_{\mu} g &=  e^{-i y(x)^a \bar{P}_a} \left( \partial_{\mu} + i \breve{A}_{\mu}^{} Q + i \breve{e}_{\mu}^{\;\; a} P_a + \frac{i}{2} \breve{\omega}_{\mu}^{ab} J_{ab} \right) e^{iy(x)^a \bar{P}_a} \\
&= \partial_{\mu} +  i A_{\mu} Q + i e_{\mu}^{\;\;a} P_a +  \frac{i}{2} \omega_{\mu}^{ab} J_{ab},
\label{eq:maurercartangravity}
\end{split}
\end{flalign}

\noindent where we have used the commutation relation rules of the symmetries (See section \ref{subsec:symmetries}) to obtain

\begin{flalign}
e_{\mu}^{\;\; a} &= \breve{e}_{\mu}^{a} + \partial_{\mu} y^a + \breve{\omega}_{\mu \;  b}^{a} y^b, \label{eq:efield}\\
A_{\mu} &=   \breve{A}_{\mu} + \partial_{\mu} \xi (y), \\
\omega^{ab}_{\mu} &= \breve{\omega}^{ab}_{\mu}.
\end{flalign} 

\noindent The field, $e^{a}_{\mu}$, is the vierbein, that appears in the tetrad formalism, which defines the metric as $g_{\mu \nu} = \eta_{ab} e_{\mu}^a e_{\nu}^b$. It can be used to build up the invariant element, $\mathrm{d}^4 x \, \mathrm{det}\,e$, as well as to change from orthogonal frame, i.e. $A_{\mu} = e_{\mu}^{\; b} A_{b}$.  The field, $\omega_{\mu}^{ \; ab}$, is known as the spin connection, and it is named given its transformation properties, which transforms as a gauge field \cite{Penco:2020kvy} (See section \ref{subsec:symmetries}).

Following the coset recipe, we can introduce the covariant derivative for matter fields by using the coefficients from the unbroken Lorentz generators, which transform in a linear representation under Lorentz transformations. The covariant derivative  for matter fields reads

\begin{equation}
\nabla^{g}_{a} = (e^{-1})_a^{\;\;\mu}(\partial_{\mu}  + \frac{i}{2} \omega_{\mu}^{bc} J_{bc}),
\label{eq:cdg}
\end{equation}

\noindent where the upper index $g$ denotes gravity, which transforms as, $\nabla_a^{g} = \Lambda_{a}^{\;\;b} \tilde{\nabla}_{b}^{g}$, under Lorentz transformations.  In a similar manner we can obtain the covariant derivative for charged fields, which transform in a linear representation under U(1) transformations, by using the coefficients from the unbroken U(1) generators. The covariant derivative for charged fields reads

\begin{equation}
\nabla^{q}_{a} = (e^{-1})_a^{\;\;\mu}(\partial_{\mu} + iA_{\mu}Q),
\label{eq:cdq}
\end{equation}

\noindent which transforms in the same way under Lorentz transformations as eq. (\ref{eq:cdg}). The only required ingredients to describe the non-linear realizations of translations and the local transformations of the Poincaré and U(1) group,  are the covariant derivatives and the vierbein \cite{Delacretaz:2014oxa}. Having defined these elements, we proceed to extract building blocks from the curvature invariants.

The curvature invariants can be obtained from the covariant version of the commutator of the covariant derivative that appear in the Maurer-Cartan form, 

\begin{flalign}
\begin{split}
g^{-1} [D_{\mu}, D_{\nu}] g &= i F_{\mu \nu} Q +  i T^{a}_{\mu \nu} P_a + \frac{i}{2} R^{ab}_{\mu \nu} J_{ab},\\
&= i\left(\partial_{\mu} A^{}_{\nu} - \partial_{\nu} A^{}_{\mu} \right)Q + i \left(\partial_{\mu} e^{a}_{\nu} - \partial_{\nu} e^{a}_{\mu}  + e^{}_{\mu b} \omega^{ab}_{\nu} - e^{}_{\nu b} \omega^{ab}_{\mu}\right)P_a\\
&+ \frac{i}{2} \left( \partial_{\mu} \omega^{ab}_{\nu} -\partial_{\nu} \omega_{\mu}^{ab} + \omega^{a}_{\mu c}\omega^{cb}_{\nu} - \omega^{a}_{\nu c} \omega^{cb}_{\mu} \right) J_{ab},
\label{eq:invariantgravity}
\end{split}
\end{flalign}

\noindent with $T^{a}_{\mu \nu} = \breve{T}^{a}_{\mu \nu} + \breve{R}^{ab}_{\mu \nu} y_b$, and $R^{ab}_{\mu \nu} = \breve{R}^{ab}_{\mu \nu}$, the covariant torsion and Riemann tensor respectively. The covariant quantities have been defined in this way, from,  $[D_{\mu}, D_{\nu}] =  i \breve{F}_{\mu \nu}^{} Q +  i \breve{T}^{a}_{\mu\nu} P_a + \frac{i}{2} \breve{R}^{ab}_{\mu \nu } J_{ab}$, in eq. (\ref{eq:commutatorflat}), such that by construction, $T^{a}_{\mu}$ and $R^{ab}_{\mu \nu}$ transforms independently under the local transformations \cite{Delacretaz:2014oxa}.

We are interested in a gravitational theory as General Relativity where the torsion tensor is zero. Solving for the vanishing torsion tensor, we obtain an equation for the spin connection in terms of the vierbein \cite{Delacretaz:2014oxa},

\begin{equation}
\omega_{\mu}^{ab} (e) = \frac{1}{2} \left\{ e^{\nu a} (\partial_{\mu} e_{\nu}^{\;\;b} - \partial_{\nu} e_{\mu}^{\;\;b}) + e_{\mu c} e^{\nu a} e^{\lambda b} \partial_{\lambda} e_{\nu}^{\;\;c} - (a \leftrightarrow b)\right\}.
\label{eq:spinc}
\end{equation}

\noindent From the expression for the torsion in eq. (\ref{eq:invariantgravity}), one can read the Christoffel symbols, $T_{\mu \nu}^a = \Gamma^{a}_{\mu \nu} - \Gamma_{\nu \mu}^a$. Therefore, the Christoffel symbol reads

\begin{flalign}
\Gamma_{\mu \nu}^a = \partial_{\mu} e^{\;a}_{\nu} - e_{\nu b} \omega_{\mu}^{ab},
\label{eq:cristoffele}
\end{flalign}

\noindent which can be used to express the Riemann tensor in terms of the Christoffel symbols as usual,

\begin{flalign}
	R^{c}_{\; dab} = \partial_{a} \Gamma^c_{bd} - \partial_{b} \Gamma^c_{ad} +  \Gamma^c_{fa} \Gamma^f_{db} -  \Gamma^c_{fb}  \Gamma^f_{da}.
\label{eq:riemanne}
\end{flalign}

After imposing the condition that the torsion tensor vanishes, eq. (\ref{eq:invariantgravity}), reads

\begin{flalign}
\begin{split}
g^{-1} [D_{\mu}, D_{\nu}] g = i F_{\mu \nu} Q + \frac{i}{2} R^{ab}_{\mu \nu} J_{ab},
\end{split}
\end{flalign}

\noindent Therefore, we can use the tensors, $F_{\mu \nu}$ and $R^{ab}_{\mu \nu}$, to build up a Lagrangian that matches to the one of the theory of General Relativity, as well as electrodynamics in curved spacetime. Such action reads:

\begin{equation}
\mathcal{S} =  \int \mathrm{det} \; e \; \mathrm{d}^4 x \left\{ -\frac{1}{4 \mu_0} F_{a b} F^{a b} +  \frac{1}{16 \pi G} R  + .\;.\;.    \right\},
\label{eq:generalactionelectrograv}
\end{equation}

\noindent  with, $R = R^{ab}_{\; \; \; ab} = e^{\mu}_{\;a} e^{\nu}_{\;b} R^{ab}_{\; \; \; \; \mu \nu}$, the Ricci scalar, which is the low energy term of the theory of gravity. The second term is the well known Einstein-Hilbert action,  and the first term is Maxwell's action in curved spacetime, with $\mu_0$, the magnetic permeability of vacuum. The coefficients of eq. (\ref{eq:generalactionelectrograv}), have been matched from the known theory, which allows us to obtain the Einstein-Maxwell action in the vierbein formalism. The ellipsis stands for higher order terms from a more fundamental theory of gravity. 

One can easily obtain the well known gravitational action by changing to the spacetime indices:

\begin{equation}
\mathcal{S} =  \int \sqrt{-g} \; \mathrm{d}^4 x    \frac{1}{16 \pi G } R  ,
\label{eq:generalactiongravEins}
\end{equation}

\noindent with $R = R^{\mu \nu}_{\;\;\; \mu \nu} = e^{\mu}_{\;a} e^{\nu}_{\;b} R^{ab}_{\; \; \; \mu \nu} $ , and where we have used $g_{\mu \nu} = e_{\mu}^a e_{\nu}^b \eta_{ab}$. 


\subsection{Charged Spinning Compact Objects}

The description of extended objects in the EFT framework was first introduced in \cite{Goldberger:2004jt, Goldberger:2005cd} for the non-spinning case. Then, spinning extended objects was introduce in \cite{Porto:2005ac}, and later in \cite{Levi:2008nh}, theories whose constructions differ. Finally, the effective theory for spinning extended objects derived using the coset construction was introduced in \cite{Delacretaz:2014oxa}. Although in \cite{Goldberger:2005cd}, BHs electrodynamics is introduced, it was until \cite{Patil:2020dme} that charge was considered in the EFT for extended objects to obtain the dynamics. In the following, we will use the coset construction and extend the work on spinning extended objects in \cite{Delacretaz:2014oxa}, to include the U(1) symmetry in order to describe charged spinning compact objects.

\subsubsection*{Charged Spinning Particles}

In the previous section we constructed an invariant action for a charged extended object without spin in flat spacetime, by identifying the symmetry breaking pattern of a charged point particle, which breaks spatial translations and boosts. The state of a charged point particle is an eigenstate of the charge and does not break the U(1) symmetry. Now, in the curved spacetime case, we consider spinning objects which break the full Poincaré group, such that the system is described by the charged point particle with the additional breaking of rotations. BHs and NSs have their own internal symmetry, which characterizes the low energy dynamics of the effective theory. The additional symmetry group for a compact object is the internal symmetry group, S $\subseteq$ SO(3), so that G = U(1) $\times$ ISO(3,1) $\times $ S, with G the full symmetry before being broken \cite{Delacretaz:2014oxa}. In principle, S, can be a discrete group, but with the appropriate choice of the coset parametrization, we can treat in the same way both continuous and discrete symmetries.

By considering the comoving frame, the group G is broken into a linear combination of  internal rotations, $S_{ij}$, and spatial rotations, $J_{ij}$, such that the symmetry breaking pattern for the charged spinning point particle reads,

\begin{flalign}
\begin{split}
\mathrm{Unbroken \; generators} &=
\begin{cases}
&\bar{P}_0 = P_0 + Q \;\;\;\;\;\;\;\;\;\, \mathrm{time \; translations},\\
&\bar{J}_{ij} \;\;\;\;\;\;\;\;\;\; \;\;\;\;\;\;\;\;\;\; \;\;\;\;\; \mathrm{\, internal \; and \; spacetime \; rotations.}	
\end{cases}\\
\mathrm{Broken \; generators} &= 
\begin{cases}
&P_i \;\;\;\;\;\;\;\;\;\;\;\;\;\;\;\;\;\;\;\;\;\;\;\;\;\; \mathrm{spatial \; translations} , \\ 
&J_{ab}\;\;\;\;\;\;\;\;\;\;\;\;\;\;\; \;\;\;\;\;\;\;\;\;\; \mathrm{boosts \; and \; rotations},
\end{cases}
\end{split}
\end{flalign}

\noindent with, $\bar{J}_{ij}$, the sum of the internal and spacetime rotations \cite{Delacretaz:2014oxa}, and where we have considered a spherical star at rest, such that, $S_{ij}$, are the generators of the internal $SO(3)$ group, and $ \bar{J}_{ij} = S_{ij} + J_{ij}$. Translations are non-linearly realized the local Poincaré and U(1) transformations are consider to take place along the fiber. Given that Lorentz transformations can be parametrized as the matrix product of a boost and a rotation, the coset parametrization reads

\begin{equation}
g = e^{i y^a \bar{P}_a} e^{i \alpha_{ab} J^{ab}/2} = e^{i y^a \bar{P}_a} e^{i \eta^j J_{0j}} e^{i \xi_{ij} J_{ij}/2} = e^{i y^a \bar{P}_a} \bar{g}^{},
\label{eq:cosetcspp}
\end{equation}

\noindent which implies a one to one correspondence between the Goldstone fields, $\alpha_{ab}$ and $\eta_i$, $\xi_{ij}$ \cite{Delacretaz:2014oxa}. By parametrizing the coset with the generators of the broken spatial rotations,  the Maurer-Cartan form  can be computed without the need to specify the explicit unbroken generators of rotations, $\bar{J}_{ij}$.

We proceed as before to identify the relevant degrees of freedom from the projected Maurer-Cartan form to the worldline of the object,

\begin{flalign}
\begin{split}
\dot{x}^{\mu} g^{-1} D_{\mu} g =& \dot{x}^{\mu} g^{-1} (\partial_{\mu} + i \breve{A}_{\mu} Q + i \breve{e}_{\mu}^{\;\; a} P_a + i \breve{\omega}_{\mu}^{\;\; ab} J_{ab} ) g \\
=& \dot{x}^{\mu} \bar{g}^{-1}(\partial_{\mu} +  i A_{\mu} Q + i e_{\mu}^{\;\;a} P_a +  \frac{i}{2} \omega_{\mu}^{ab} J_{ab}) \bar{g} \\
=& ie( P_0 + AQ + \nabla \pi^i P_i + \frac{1}{2} \nabla \alpha_{cd} J^{cd}).
\end{split}
\end{flalign}

\noindent The building blocks of the low energy dynamics are,

\begin{flalign}
e &= \dot{x}^{\mu}  e_{\mu}^{\; \; a} \Lambda_a^{\;\; 0} \\
A &= e^{-1} \dot{x}^{\mu} A_{\nu}^{}  \Lambda_{\;\; \mu}^{\nu}   \\
\nabla \pi^i &= e^{-1} \dot{x}^{\mu} e_{\mu}^{\; \; a} \Lambda_a^{\;\; i}  \\
\nabla \alpha^{ab} &= e^{-1} \left( \Lambda_c^{\;\; a} \dot{\Lambda}^{cb} + \dot{x}^{\nu} \omega_{\nu}^{cd} \Lambda_c^{\;\; a} \Lambda_d^{\;\; b} \right)  
\label{eq:spinbuilding}
\end{flalign}

\noindent  with the $\Lambda$'s, the Lorentz transformations parametrized by $\alpha$, or equivalently by $\eta$ and $\xi$. As one can observe, one of the advantages of using the parametrization (\ref{eq:cosetcspp}), is that no connection proportional to $\bar{J}$ appears, which makes the building blocks independent of the residual symmetry group \cite{Delacretaz:2014oxa}. The residual symmetry, SO(3), requires all spatial indices to be contracted in an SO(3) invariant manner.

From the coset perspective, we can remove some of the Goldstones using the inverse Higgs constraint \cite{Ivanov:1975zq}, given that the commutator between the unbroken time translations and boosts, gives spatial translations, $[K^i, P^0] = i P^i$. Thus, we set to zero the covariant derivative of the Goldstone, 

\begin{equation}
\nabla \pi^i = e^{-1} (\dot{x}^{\nu} e_{\nu}^{0} \Lambda_{0}^{\; \; i} + \dot{x}^{\nu} e_{\nu}^{j} \Lambda_{j}^{\; \; i}) = 0,
\label{eq:higgs-constraint}
\end{equation}

\noindent and solve this constraint, to express the boost as a function of velocities  \cite{Delacretaz:2014oxa},

\begin{equation}
\beta_i \equiv \frac{\eta_i}{\eta} \tanh \eta = \frac{\dot{x}^{\nu} e_{\nu}^{\; \; i}}{\dot{x}^{\nu} e_{\nu}^{\; \; 0}}
\label{eq:beta}
\end{equation}

\noindent where, $\eta$, is identified as the rapidity (See appendix \ref{subsec:symmetries}). To obtain the physical interpretation of last equation, we rewrite the building block, $e$, such that

\begin{flalign}
\begin{split}
e =& \sqrt{(e \nabla \pi^i)^2 - (\dot{x}^{\nu} e_{\nu}^{\;\; a}\Lambda_{a}^{\;\; c} \dot{x}^{\mu} e_{\mu}^{b} \Lambda_{bc})}\\
=& \sqrt{-\eta_{ab} e_{\nu}^{\;\; a} e_{\mu}^{\;\; b} \dot{x}^{\nu} \dot{x}^{\mu}} = \sqrt{- g_{\mu \nu} \dot{x}^{\mu} \dot{x}_{\mu}} = \frac{\mathrm{d} \tau}{\mathrm{d} \sigma},
\end{split} 
\label{eq:E}
\end{flalign}

\noindent with, $\sigma$, the worldline parameter that traces out the trajectory of the particle. To obtain last equation we have imposed the inverse Higgs constraint, and the property of the boost matrices, $\Lambda_{a}^{\;\; b} \Lambda^{a}_{\;\; c} = \delta^{b}_{\;\;c}$. 

Therefore, with eq. (\ref{eq:E}), we rewrite the constraint, eq. (\ref{eq:higgs-constraint}), in a way that makes manifest its physical interpretation \cite{Delacretaz:2014oxa}. For rotations, we can write, $\Lambda^0_{\;\;a} (\xi) = \delta^0_a$ and $\Lambda^i_{\; j} (\xi) = \mathcal{R}^i_{\;\;j} (\xi)$, with $\mathcal{R}(\xi)$ an $SO(3)$ matrix, such that the constraint (\ref{eq:higgs-constraint}) now reads

\begin{equation}
u^a \Lambda_a^{\;\; i} (\eta) \mathcal{R}_{i}^{\;\; j} (\xi) = 0.
\end{equation}

\noindent with, $u^a \equiv e_{\mu}^{\; \; a} \partial_{\tau} x^{\mu}$, the Lorentz velocity measured in the local inertial frame defined by the vierbein. Given that the matrix $\mathcal{R}^{\;\;i}_j (\xi)$ is invertible, we obtain 

\begin{equation}
u^a \Lambda_a^{\;\; i} (\eta) = 0.
\end{equation}

\noindent  These quantities have a clear geometrical interpretation: the set of local Lorentz vectors, 

\begin{equation}
{\hat{n}^a_{\;\; (0)} \equiv u^a  = \Lambda^a_{\;\; 0} (\eta) \;, \;\;\;\; \hat{n}^{a}_{(i)} \equiv \Lambda^a_{\;\; i} (\eta)   },
\label{eq:neta}
\end{equation}

\noindent that define an orthonormal local basis with respect to the local flat metric, $\eta_{a b}$, in the frame that is moving with the particle \cite{Delacretaz:2014oxa}. Orthogonality is obtained from the boost matrices, $\Lambda_a^{\;\; b} \Lambda^a_{\;\; c} = \delta^b_c$. This means that, the $\hat{n}^a_{(b)}$, is a set of vierbeins that defines the local inertial frame on the particle worldline. One can also define the orthonormal basis in terms of the spacetime vectors, $\hat{n}^{\mu}_{\;(b)} \equiv e^{\mu}_{a} \hat{n}^a_{(b)}$, with respect to the full metric, $g_{\mu \nu}$.   

Moreover, an additional set of orthonormal vectors is obtained \cite{Delacretaz:2014oxa},

\begin{equation}
\hat{m}^b_{\;\; (a)} \equiv \Lambda^b_{\; \; a} (\alpha) = \Lambda^b_{\; \; c} (\eta) \mathcal{R}^c_{\; \; a} (\xi),
\label{eq:orthom}
\end{equation}

\noindent with the zeroth vector, $\hat{m}^b_{\;\;(0)} = \hat{n}^b_{\;\;(0)} = u^b$, the velocity of the compact object in the proper frame. The rest of the vectors differs by a rotation, $\mathcal{R} (\xi)$, compared to (\ref{eq:neta}), from which one can observe that the set of vectors, ${\hat{m}^b_{\;\; (a)}}$, contain information about the rotation, parametrized by the three degrees of freedom of $\xi$.

With these definitions, the covariant derivatives, $\nabla \alpha^{0i}$, can be expressed as

\begin{equation}
 \nabla \alpha^{0i}  = \mathcal{R}_{j}^{\; \; i} (\xi) \Lambda_{a}^{\;\; (j)} (\eta) (\partial_{\tau} u^a + u^{\mu} \omega_{\mu\;\;c}^{\;\;a} u^c) = \mathcal{R}_{j}^{\; \; i} (\xi) \Lambda_a^{\; (j)}(\eta) a^{a} = a^i ,
 \label{eq:temporalspinalpha}
\end{equation}

\noindent which is a rotated version of the acceleration projected into the orthonormal basis defined by the $\hat{n}$'s basis in the proper frame of the particle.  As in the case of charged spheres, in the absence of external forces, this building block is zero. Nevertheless, it must be considered to build up invariant terms when an external force, such as the one from an external gravitating object, or form the charge of another object, is strong enough to make this building block relevant to the dynamics. 

The rest of the covariant derivatives for the Goldstone, reads

\begin{equation}
\nabla \alpha^{ij} = \Lambda^{\;\;i}_{k}  (\eta^{kl} \partial_{\tau} + \partial_{\tau} x^{\mu} \omega_{\mu}^{kl}) \Lambda_{l}^{\;\; j} = \Omega^{ij},
\end{equation}

\noindent which is the relativistic angular velocity of the object in the proper frame. From the covariant quantity, $\Omega^{ij}$, it is possible to define the angular velocity vector with the epsilon tensor, which reads 

\begin{equation}
\Omega_i = - \frac{1}{2} \epsilon_{ijk} \Lambda_{a}^{\;\; j}(\eta^{ab} \partial_{\tau} + \partial_{\tau} x^{\mu} \omega_{\mu}^{ab})\Lambda_{b}^{\;\;k}.
\label{eq:spinepsilon}
\end{equation}

Thus, we are left with the building blocks

\begin{flalign}
e &= \mathrm{d} \tau/ \mathrm{d} \sigma, \\
A &= u^{\mu} A_{\mu}^{},  \\
\nabla \alpha^{0i} &= a^i \\
\nabla \alpha^{ij} &= \Omega^{ij}, 
\end{flalign}

\noindent with $A_{\mu}^{} = A_{\nu}^{}  \Lambda_{\;\; \mu}^{\nu}$. We consider the blocks $A$ and $e$, to derive the effective action of a charged point particle in curved space-time. The action takes the form

\begin{equation}
\mathcal{S} = \int \mathrm{d} \sigma e ( -n_E  + n_A A ) = \int \mathrm{d} \tau ( -mc^2  + qu^{a}  A_{a} ),
\label{eq:ppchg}
\end{equation} 

\noindent where we have matched the coefficient, $n_E = mc^2$, from the action of a point particle \cite{Goldberger:2004jt} with mass, $m$, and the coefficient, $n_A = q$, from the action of a charged point particle \cite{Goldberger:2004jt,Patil:2020dme} with net charge, $q$. 

Now we include the building block, $\nabla \alpha^{ab}$, neglecting charge for simplicity. The leading order action reads \cite{Delacretaz:2014oxa},

\begin{equation}
\mathcal{S} = \int \mathrm{d} \sigma e \left(-mc^2 + n_{\alpha} \nabla \alpha_{ij} \nabla \alpha^{ij} + .\;.\;. \right) = \int \mathrm{d} \sigma e \left(-mc^2 + n_{\Omega} \Omega_{ij} \Omega^{ij}  \right) , 
\end{equation}
\noindent where a term linear in $\Omega$, has been discarded by time reversal symmetry, and we have considered spherical objects at rest. The ellipses denotes higher order corrections made out of $\nabla \alpha^{ab}$. In the mechanics of rotational dynamics, to characterize a rigid sphere only two parameters are needed, which is the mass, $m$, and moment of inertia, $I$. By comparing our action to the one of a relativistic spinning point particle \cite{HANSON1974498,Porto:2005ac}, we match the coefficient, $n_{\Omega} = I/4$, to obtain the relativistic action for spinning particles in curved space-time \cite{Delacretaz:2014oxa, Endlich:2015mke},

\begin{equation}
\mathcal{S}  =  \int \mathrm{d} \tau \left\{ -m + \frac{I}{4} \Omega_{ab} \Omega^{ab} + .\;.\;.  \right\},
\end{equation}

\noindent where we have defined, $  \Omega_{0i} = 0$, as in \cite{Endlich:2015mke}. A higher order correction made out of the Goldstone building block,

\begin{flalign}
\mathcal{S}  =  \int \mathrm{d} \tau c_{\alpha,u} \nabla \alpha_{ac} \nabla \alpha^{cb} \frac{u^a u_b}{u^2}  = \int \mathrm{d} \tau  c_{\Omega,a}  \Omega_{ij} a^i \frac{u^j u_0}{u^2} +  \;.\;.\;.\;  .
\label{eq:spincorrection}
\end{flalign}

We can connect our action to the one used to obtain the PN expansion for spinning objects \cite{Porto:2008jj, Levi:2015msa}, with the introduction of the relativistic spin degree of freedom \cite{Porto:2005ac,Levi:2015msa,Steinhoff:2021dsn}

\begin{flalign}
S^{ab} = 2 \frac{\partial \mathcal{L}}{\partial \Omega_{ab}},
\end{flalign}

\noindent which is the conjugate variable of the angular velocity, with associated spin tensor $S^{ab} = \epsilon^{ab}_{\;\;\;c}S^c$. Then, by considering the action that describes a spinning extended object in the proper frame, 

\begin{equation}
\mathcal{S}  =  \int \mathrm{d} \tau \left\{ -mc^2 + \frac{I}{4} \Omega_{ab} \Omega^{ab} +  c_{\Omega,a} \Omega_{ab} a^a \frac{u^b u_0}{u^2} + .\;.\;.  \right\},
\end{equation}

\noindent we Legendre transform it to write it down in terms of the spin $S^{ab}$, and transform it to the lab frame, with 

\begin{flalign}
\tilde{\Omega}^{ab} = \Lambda^{a}_{\;\;c} \Lambda^{b}_{\;\;d} \Omega^{cd} & = \Lambda^{a}_{\;\;c} \Lambda^{b}_{\;\;d} [ \Lambda_{e}^{\;\;c} (\eta^{ef} \partial_{\tau} + \partial_{\tau} x^{\mu} \omega_{\mu}^{\; \; ef})\Lambda_{f}^{\;\;d}  ]\\
&= - \Lambda^{a}_{\;\;c} \partial_{\tau} \Lambda^{bc} + \partial_{\tau} x^{\mu} \omega_{\mu}^{\;\;ab},
\label{eq:angularvel2}    
\end{flalign}

\noindent and $\tilde{S}^{ab} = \Lambda^{a}_{\;\;c} \Lambda^{b}_{\;\;d} S^{cd}$. In the lab frame, in which the PN expansion is computed, we set $u^i = v^i$ with $u^0 = 1$, and $\sigma = t$, such that we obtain

\begin{flalign}
\mathcal{L} = -mc^2 + \frac{1}{2} S_{ab} \Omega^{ab} + \frac{c_{\Omega,a}}{I} S_{ab} \frac{ a^{a} u^b}{ u^2} - \frac{1}{4I} S_{ab}S^{ab},  
\label{eq:lomega2}
\end{flalign}

\noindent with the expected acceleration correction in \cite{Levi:2015msa} when setting $c_{\Omega,a} =I$.

Since $u^a \Lambda_a^{\;\; i} = 0$, the tensors, $\nabla \alpha^{ab}$ and $S^{ab}$, are orthogonal to the four velocity. Therefore, we can obtain a constraint on the angular velocity, 
 
\begin{flalign}
u_a \nabla \alpha^{ab} = u^{\mu} \nabla_{\mu} u^{b} + u_a \Omega^{ab} = 0,   
\end{flalign}

\noindent as well as on the spin, 

\begin{flalign}
u_a S^{ab}  = \sqrt{u^2} S^{0b} + u_i S^{ib} =  0.   
\end{flalign}

\noindent The latter, known as the spin supplementary condition, is equivalent to the relativistic Price-Newton-Wigner spin supplementary condition \cite{Steinhoff:2015ksa,Levi:2015msa}, while the former is the constraint on the angular velocity derived in \cite{Endlich:2015mke}.

With all of our derived invariant operators, we write down the effective action of a charged spinning point particle in the particle's rest frame, 

\begin{equation}
\mathcal{S}  =  \int \mathrm{d} \tau \left\{ -mc^2 + qu^{a} A_{b} + \frac{I}{4} \Omega_{ab} \Omega^{ab} +  I  \Omega_{ab} a^a \frac{u^b u_0}{u^2} + .\;.\;.  \right\}.
\label{eq:chspp}
\end{equation}

\subsubsection*{Size Effects}

Size effects in the EFT for extended objects was introduced in \cite{Goldberger:2004jt}, which can be systematically taken into account by building invariant operators made up of the Weyl curvature tensor, $W_{abcd}$, that is obtained from the Riemann tensor, $R_{abcd}$, by subtracting out various traces. By defining the transformation \cite{Delacretaz:2014oxa}

\begin{equation}
R \equiv g^{}_{L} \; \tilde{R},
\end{equation}  

\noindent with, $g_{L}$, the Lorentz part of the parametrization eq. (\ref{eq:cosetcspp}), the Riemann tensor transforms linearly under Lorentz transformations as expected. The explicit transformation reads

\begin{equation}
R_{abcd} = \Lambda_{a}^{\;\; e} \Lambda_{b}^{\;\; f} \Lambda_{c}^{\;\; g} \Lambda_{d}^{\;\; h} \tilde{R}_{efgh},
\label{eq:Riemannproper}
\end{equation}

\noindent with, $\tilde{R}_{abcd}$, the Riemann tensor in the local rest frame of the object.  Having the correct transformations, we define the Weyl tensor as usual,  

\begin{equation}
W_{abcd} = R_{abcd} + \frac{1}{2} (R_{ad}g_{bc} - R_{ac}g_{bd} + R_{bc}g_{ad} - R_{bd}g_{ac}) + \frac{1}{6} R (g_{ac} g_{bd} - g_{ad} g_{bc}),
\end{equation}

\noindent which have the physical content \cite{Goldberger:2004jt}. The Weyl tensor measures the curvature of the spacetime and contains the tidal force exerted on an extended particle that is moving along the worldline, taking into account for how the shape of the body is distorted.  It transforms as eq. (\ref{eq:Riemannproper}). 

Furthermore, we can also use the electromagnetic tensor to build invariant operators that take into account for the polarizability of the object. Following the above discussion, we define its transformation rule,

\begin{equation}
F \equiv g^{}_{L} \; \tilde{F}.
\end{equation}  

\noindent  Having defined the correct transformation rules for the Riemman and the electromagnetic tensor, we can now proceed to form rotationally invariant objects.

We form all leading order invariant operators that contribute to the dynamics, by combining all of our covariant quantities: $W^{abcd}$, $F^{ab}$,  $\nabla \alpha^{ab}$ and $u^a$, in all possible ways allowed by the symmetries. In particular, for the electromagnetic and Weyl tensor, the building blocks are the electric like parity tensors, $E^{}_{a} = F_{ab} u^b$, and  $E^{}_{ab} = W_{acbd} u^c u^d$, respectively  \cite{Goldberger:2005cd}. By considering the electromagnetic dipolar and gravitational quadrupolar moments, we build the following leading order relevant operators for finite-size effects:

\begin{flalign}
\begin{split}
\tilde{\mathcal{O}}(u^a,\Omega^{a},E^{ab},E^{a}) &=
\begin{cases}
& E^{ab} E_{ab}  \;\;\;\;\;\;\;\;\;\;\;\;\;\;\;\;\;\;\; \mathrm{Gravity},\\
& \Omega^a \Omega^b E_{ab}  \,\;\;\;\;\;\;\;\;\;\;\;\;\;\;\,  \mathrm{Spin-gravity},\\
& E_{}^{a} E_{a}  \;\;\;\;\;\;\;\;\;\;\,\;\;\;\;\;\;\;\;\; \;\; \mathrm{Electromagnetic,}\\
& \Omega^a E_a \;\;\;\;\;\;\;\;\;\;\;\;\;\;\;\;\;\;\;\;\; \mathrm{Spin - electro.}\\
\end{cases}
\end{split}
\label{eq:operators}
\end{flalign}

One can consider the magnetic parity operators as well, $B_{ab}^{} = (1/2) \epsilon_{cdea}W^{cd}_{\;\;\;\;fb} u^e u^f $ and $B_a^{} = \epsilon_{abcd} F^{bc} u^d$ \cite{Goldberger:2005cd}, for the gravitational and electromagnetic case respectively, which are subleading with respect to the electric parity terms (at least for the gravitational case), therefore restricting our discussion to the electric parity action. The leading order magnetic parity operators can be build in analog to (\ref{eq:operators}) i.e. $B^{ab} B_{ab}$, $B^{a} B_{a}$, etc. 

Higher order terms to the ones shown in eqs. (\ref{eq:operators}), can be built from the derived building blocks and the covariant derivatives, eqs. (\ref{eq:cdg}) and (\ref{eq:cdq}), i.e. $\nabla_{c}^g E_{ab}$ and $\nabla_{b}^q E_{a}$. Furthermore, it is worth noting that size effects can be seen as encoded in a composite operator $Q^{ab}$, which we comment below.

\subsubsection*{Dissipative Effects}

Dissipation, due to the internal structure of an extended object, was introduced in EFT description in \cite{Goldberger:2005cd}, where the existence of gapless modes that are localized on the worldline of the particle take into account for the energy and momentum loss from the interaction with external sources. Dissipative effects for slowly spinning objects were considered in \cite{Porto:2007qi, Endlich:2015mke}, and for maximally spinning in \cite{Goldberger:2020fot}.  These large number of degrees of freedom can be encoded in operators allowed by the symmetries of the object. For a compact object, the allowed operators due to its symmetries gives rise to the invariant operators \cite{Goldberger:2005cd, Goldberger:2020fot}:

\begin{flalign}
\begin{split}
\mathrm{Dissipative}\mathrm{\; operators} &=
\begin{cases}
&\mathcal{P}^a (\sigma) F_{ab}u^b \;\;\;\;\;\;\;\;\;\;\;\;\;\;\;\;\;\; \mathrm{Electro},\\
& \mathcal{D}^{ab} (\sigma) W_{acbd} u^c u^d \;\;\;\;\;\;\;\;\;  \mathrm{Gravity},
\label{eq:dissoper}
\end{cases}
\end{split}
\end{flalign}

\noindent with $\mathcal{P}(\tau)$ and $\mathcal{D}(\tau)$, composite operators corresponding to the electric parity of the electromagnetic dipole and the gravitational quadrupole moment respectively, encoding the dissipative degrees of freedom. 

For a non-spinning BH, dissipation takes into account
for the absorption of electromagnetic and gravitational waves, while for a non-spinning NS, dissipative effects take into account for the energy loss during the interaction with an external source given the internal equation of state of matter. On spinning objects, the spin has a time dependence between the object and its environment which generates dissipative effects. The operators in eq. (\ref{eq:dissoper}), take into account for the spin dissipative effects as well. The coefficients do not appear explicitly in the Lagrangian, but they are encoded in the dynamical moments, $\mathcal{P}$ and $\mathcal{D}$, which are dependent on the internal degrees of freedom of the compact object in an unspecified way, but which explicit form is not necessary to obtain the dynamics \cite{Goldberger:2005cd, Goldberger:2020fot}. The dynamics of the system containing the dissipative degrees of freedom can be obtained using the in-in closed time path \cite{Jordan:1986ug}, a formalism that allows us to treat dissipative effects in a time asymmetric approach \cite{Goldberger:2005cd}.

The expectation values of these operators, $\braket{\mathcal{P}^a(\tau)},\,\braket{\mathcal{D}^{ab}(\tau)}$, are defined through the in-in path integral, which is the expectation value in the initial state of the internal degrees of freedom, and which in general is a functional of the building blocks, $E^{a}$ and  $E^{ab}$, respectively \cite{Goldberger:2020fot}. In the gravitational case, by considering the linear response in a weak external field, the in-in expectation values implies the form of the expectation value \cite{Goldberger:2020fot}

\begin{flalign}
\braket{\mathcal{D}^{ab} (\tau)} = \int \mathrm{d}\tau' G^{ab,cd}_{R} (\tau - \tau') E_{cd} (\tau') + O\left(E^2\right),
\end{flalign}

\noindent where the expectation values of the retarded Green's function,

\begin{flalign}
G^{ab,cd}_{R} (\tau - \tau') = i \theta (\tau -\tau') \braket{[\mathcal{D}^{ab}(\tau),\mathcal{D}^{cd}_{}(\tau')]},
\label{eq:greensr}
\end{flalign}

\noindent are obtained at the initial state of the interaction where the external field is zero. By considering low frequencies, from which we assume that the degrees of freedom from the operator, $\mathcal{D}^{ab}$,  are near equilibrium, the time ordered two point correlation function imply that the Fourier transform $G_R$ must be an odd, analytic function of the frequency, $\omega > 0$ \cite{Goldberger:2005cd}. Therefore, the retarded correlation function, $\tilde{G}_R$, reads

\begin{equation}
    G^{ab,cd}_R (\omega) \simeq i c_g \omega \left( \delta^{ac} \delta^{bd} + \delta^{ad} \delta^{cb} - \frac{2}{3} \delta^{ab} \delta^{cd} \right),
    \label{eq:retarded}
\end{equation}

\noindent with the coefficient for dissipative effects, $c_g \geq 0$. Note that, in contrast to \cite{Goldberger:2005cd}, we have absorbed the $1/2$ factor appearing in front of (\ref{eq:retarded}) into the dissipative coefficient. 

By considering the response of the interaction to be nearly instantaneous, the operator due to gravitational dissipative effects takes the form \cite{Goldberger:2005cd,Goldberger:2020fot}

\begin{flalign}
\braket{\mathcal{D}_{ab} (\tau)} \simeq   ic_{g} \frac{\mathrm{d}}{\mathrm{d}\tau}  E_{ab} + \,.\;\;.\;\;.\,.
\label{eq:response}
\end{flalign}

\noindent In general, one can use the dynamical composite operator, $Q_{ab} (\tau)$, to account for the tidal response function, for which the above reasoning leads to \cite{Goldberger:2005cd,Goldberger:2020fot}

\begin{flalign}
\braket{Q_{ab} (\tau)} \simeq n_{g} E_{ab} +  ic_{g} \frac{\mathrm{d}}{\mathrm{d}\tau}  E_{ab} + n_{g}^{'} \frac{\mathrm{d}^2}{\mathrm{d}\tau^2}  E_{ab} + \,.\;\;.\;\;.\,.
\label{eq:responseQ}
\end{flalign} 

\noindent The first term in last equation is the static quadrupolar tidal effect considered above, while the third term is a dynamical tidal effect in the quasi-static limit. The second term, which is the imaginary part of the response function, takes into account for the dissipative degrees of freedom.  In our work we consider the static tidal response only, and use the operator, $\mathcal{D}_{ab}$, to contain only the dissipative degrees of freedom. Although in the numerical simulations for NSs we use dynamical oscillations in the quasi-static limit to study its effects, the state of the art modeling of dynamical oscillations is done in a different fashion \cite{Steinhoff:2021dsn}, which is beyond the scope of this work. 

On the electromagnetic side, an analog procedure can be taken, for which retarded correlation function reads \cite{Goldberger:2005cd}

\begin{equation}
    G^{ab}_R (\omega) \simeq i c_q \delta^{ab} \omega,
    \label{eq:retardedq}
\end{equation}

\noindent with the coefficient, $c_q \geq 0$. Therefore, the operator for electromagnetic dissipative effects reads

\begin{flalign}
\braket{ P_{a} (\tau)}  \simeq  ic_{q} \frac{\mathrm{d}}{\mathrm{d}\tau}E_{a} + \,.\;\;.\;\;.\,.
\label{eq:responseP}
\end{flalign} 

\subsection{The Effective Action}

Gathering all our results,  we construct the most general, leading order and electric like parity, effective action for a compact object in the theory of General Relativity, 

\begin{flalign}
\begin{split}
\mathcal{S}_{eff} = \int \mathrm{d}\tau  &\left\{  -mc^2 + qu^{a} A_{a} + \frac{I}{4} \Omega_{ab} \Omega^{ab} +  I  \Omega_{ab} a^b \frac{u^a u_0}{u^2}  \right. \\
&\;\;+ n_{q,\Omega} \Omega^a E_a +  n_{g,\Omega} \Omega^a \Omega^b E_{ab} + n_{q} E^{a} E_{a}  \\
& \left. \;\; +  n_{g} E^{ab} E_{ab} +   \mathcal{P}^a E_{a}  +  \mathcal{D}^{ab} E_{ab} + \;  .\;.\;. \; \right\}   + \; \mathcal{S}_{0}, \\
\end{split}
\label{eq:effectivetheory}
\end{flalign}

\noindent with the electric parity tensor, $E^{}_{ab} = W_{acbd} u^c u^d$, and $E^{}_{a} = F_{ab} u^b$, that corresponds to the gravitational quadrupole and electromagnetic dipole moment respectively. The interaction action,

\begin{flalign}
\mathcal{S}_0 = \int \mathrm{det} \; e \; \mathrm{d}^4 x \left\{ -\frac{1}{4 \mu_0} F_{a b} F^{a b} +  \frac{1}{16 \pi G} R + .\;.\;.  \right\},
\end{flalign}

\noindent is the Einstein-Maxwell action. The action describing the charged spinning compact object lives in the worldline, while $\mathcal{S}_0$ lives in the bulk.

\subsection{The Coefficients of the Effective Theory}

The coefficients of the effective theory encode the microphysics of the compact objects, which are determined through a matching procedure to the full known theory, and ultimately from GW observations. We identify them from the results in literature, without the need to do the explicit calculations here.  We already have pointed out the coefficients appearing in the action describing a charged spinning point particle with corrections due to its acceleration in eq. (\ref{eq:chspp}). The coefficient of the point particle term \cite{Goldberger:2004jt}, $c_E = mc^2$, the coefficients in the spin corrections, $c_{\Omega} = I/2$ and $c_{\Omega,u} = I$ \cite{Porto:2005ac,Levi:2015msa}, and the coefficient from the correction due to electromagnetic charge, $c_A = q$ \cite{Goldberger:2004jt,Patil:2020dme}. Now, we proceed to point out the rest of the coefficients due to the internal structure of the compact object.

We start with the coefficient due to static tidal effects, $n_{g}$, which is a coefficient that depends on the internal structure of the star through a parameter known as the Love number. Any stellar object that can be described by an equation of state of matter, can be described approximately in terms of its Love numbers. In the case of BHs, it has been found that their Love numbers vanishes $n_{g} = 0$ \cite{Binnington:2009bb}, and for  spinning BHs $n_{g,\Omega} = 1$ \cite{Chia:2020yla,Charalambous:2021mea}. Furthermore, it has been shown that for non-spinning \cite{Hui:2021vcv} and spinning BHs \cite{Charalambous:2021mea}, the coefficients due to the polarizability vanishes as well, $n_{q} = 0$ and $n_{q,\Omega} = 0$. 

On NSs, the leading order quadrupolar Love coefficient has the form \cite{Flanagan:2007ix}  

\begin{flalign}
n_{g} = \frac{2\ell^5 k_2}{3G},
\label{eq:love-coeff}
\end{flalign} 

\noindent with $k_2$, the quadrupolar dimensionless Love number, and $\ell$ the radius of the star.\footnote{We have chosen to denote the radius of the object with $\ell$ as in \cite{Endlich:2015mke}, rather than with $R$ as commonly used, given that we use $R$ for the Ricci scalar. } The Love numbers are dimensionless parameters that measure the rigidity and tidal deformability of the compact object, and varies given different equations of state of matter \cite{Yagi:2016ejg}. This number is related to other parameters of the star from the relativistic I-Love-Q relations \cite{Yagi:2016ejg,Yagi:2016bkt}, which relates the moment of inertia, $I$, the tidal deformability parameter, or Love number, $k_2$, and the quadrupole moment. These relations, which are empirically found to hold for a wide range of equations of state of matter, are only approximate relations and are not exact first principles relations. 

On the coefficient due to the coupling of spin-gravity size effects for spinning NSs, in the slow rotation limit, the relativistic coefficient, $n_{g,\Omega}$, is obtained through the Love-Q part of the I-Love-Q relations \cite{Yagi:2016bkt}. The latter coefficient in the Newtonian limit is the same as the one for static tides, $n_{g,\Omega} = n_g $ \cite{Yagi:2016bkt}.
On the charge-gravity, and charge-spin size effects,  given that charge in compact objects has been mostly neglected, the rest of the coefficients, $n_{q,\Omega}$, $n_{q}$ and $n_{q,g}$, are unknown, and are to be derived by analytical and numerical means.  

Moving on into dissipative effects, although their coefficients are not explicitly shown in the action, they are encoded in the operators, $\mathcal{P}$ and $\mathcal{D}$, as shown above. We point out the known coefficients for dissipative effects in BH interactions from the existing literature. The operator $\mathcal{D}^{ab}$, for non-spinning BHs, contains the coefficient, $c_{g}$, which encodes the capacity of the BH to absorb GWs, and which can be read off from the response function derived in \cite{Goldberger:2005cd},
\begin{equation}
    c_{g} = \frac{16}{90} \frac{G^5M^6}{c^{13}} = \frac{\ell^6_{s}}{360 G c},
    \label{eq:disscoeff}
\end{equation}

\noindent  with $M$ the mass of the BH, and $\ell_{s}$ its radius. The coefficient in eq. (\ref{eq:disscoeff}), include the factor of $1/2$ from eq. (\ref{eq:retarded}) \cite{Goldberger:2005cd}. In the same way, we can obtain the coefficient for electromagnetic dissipative effects from the composite operator, $\mathcal{P}(\tau)$ \cite{Goldberger:2005cd},

\begin{flalign}
c_q = \frac{2 \pi \ell^4_{s}}{3 \mu_0 c},
\end{flalign}

\noindent which encodes the capacity of the BH to absorb electromagnetic waves.

On the dissipative effects of rotating BHs, these effects have been considered for the slow \cite{ Endlich:2015mke, Porto:2007qi} and maximally spinning \cite{Goldberger:2020fot} cases. The coefficients can be obtained from the response function derived in \cite{Chia:2020yla} using the Teukolsky equation, for both non-spinning and spinning case. We start by reading off the coefficient for the non-spinning BH from the response function, which in the notation of \cite{Chia:2020yla}, reads

\begin{flalign}
\begin{split}
\frac{1}{2} N_2 \ell^5_{S} \mathcal{F}_{2m_s}^{Sch} =& i\frac{
\ell^5_{S} M}{90 c^3} \omega +  \mathcal{O}(\omega^3) \simeq i\frac{\ell_{S}^6}{360 Gc} \omega ,
\label{eq:chia1}
\end{split}
\end{flalign}

\noindent  where we have used the leading order mode, $l = 2$, of the angular momentum number, such that $N_2 = 1/3$, and with $m_s$, the azimuthal number, and $\ell_S$, the Schwarzschild radius. To obtain eq. (\ref{eq:chia1}) as eq. (\ref{eq:disscoeff}), we have substituted the mass of the BH in terms of the Schwarzschild radius, $M = \ell_{S} c^2 / 2G$, and considered the extra factor of $1/2$, as for eq. (\ref{eq:disscoeff}).

Now we move into the spinning case, for which the response function reads \cite{Chia:2020yla}

\begin{flalign}
\begin{split}
\mathcal{F}_{2m_a}^{I, Kerr} =&    - \frac{i}{30 Gc} \frac{is m_s}{(\ell_+ - \ell_-)} +  \frac{i}{15 c^3} \frac{\ell_+ M}{(\ell_+ - \ell_-)} \omega  + \mathcal{O}(\omega^3)  \\
\simeq&  -  \frac{i}{30 Gc}\frac{I m_s}{M c (\ell_+ - \ell_-)} \Omega + \frac{i}{15 c^3} \frac{\ell_+ M}{(\ell_+ - \ell_-)} \omega,
\end{split}
\end{flalign}

\noindent with $s = J/Mc = I \Omega/Mc $, with $J = I \Omega$, the scalar value of the angular momentum, and $\ell_+$ and $\ell_-$, the outer and inner radius of the Kerr BH.\footnote{To make the reading of the coefficients accessible, we have written the response function as in \cite{Chia:2020yla}, and then converted it to our notation. Therefore, in this case, do not confuse, $a$, with the acceleration defined in the previous section} The moment of inertia of a BH is $I = 4 GM^2/c^4$.  The, the response function with its normalization constant, reads 

\begin{flalign}
\frac{1}{2} N_2 \ell_{+}^5 \mathcal{F}_{2m_s}^{I, Kerr} \simeq &   -  \frac{i}{180} \frac{I m_s \ell_{+}^5}{M G c (\ell_+ - \ell_-)} \Omega + \frac{i}{90c^3} \frac{\ell_{+}^6 M}{(\ell_+ - \ell_-)} \omega.
\label{eq:responsespin}
\end{flalign}

\noindent Therefore, we have obtained a response function for the dissipative effects of the form,  $\braket{\mathcal{D}_{ab} (\tau,\Omega)} \propto i (c_{g,\Omega} \Omega +  c_{g} \mathrm{d}/\mathrm{d}\tau)E_{ab}$. We can identify that, for a rotating extended object, tidal dissipation arises due to two separate contributions.  The first term of eq. (\ref{eq:responsespin}), for which is nonzero even in the case of $\omega = 0$, arises given that the spin of the body has a time dependence between the object and its tidal environment. This can be seen, from the object perspective in its proper frame, as the external environment rotating with the frequency of the spin of the object.

Furthermore, now we have an expression which is explicit on the azimuthal numbers, $m_s$.  For the dominant perturbation mode, $l = 2$, then, $m_s = [-2,-1,0,1,2]$. Nevertheless, we can not identify a specific value of $m_s$ to be dominant. Therefore, it is necessary to sum over all possible values of $m_s$. This can be done by considering the size effects operators in terms of the electric and magnetic tidal moments. For instance, consider schematically the electric tidal moment, $E_{lm_s}$, for which leading order is $E_{2m_s}$. Then, we can proceed to identify the different elements of $E_{2m_s}$, given the possible values of $m_s$, to then couple each response function with its electric tidal moment. For instance, we would couple schematically, $ Q_{2m_s} \propto \mathcal{D}_{2-2} E_{2-2} + \mathcal{D}_{2-1} E_{2-1} + \mathcal{D}_{20} E_{20} + \mathcal{D}_{21} E_{21} + \mathcal{D}_{22} E_{22}$. The same reasoning applies for the magnetic parity like tidal moments. On the dissipative effects for charged spinning BHs, the coefficient, $c_{q,\Omega}$, is still unknown.

The coefficients of NSs for dissipative effects are still unknown, and must be determined from hydrodynamical simulations, analytical computations, and ultimately from observations. Nevertheless, it is worth commenting that for a stellar object that is described by an equation of state of matter, there exists the weak friction model \cite{Hut1981}, which might be applicable for NSs. For the latter, the coefficient, $c_g = \Theta n_g$  \cite{Hut1981}, with $\Theta$, being the time lag, which accounts for the tidal bulge formed in the NSs during the interaction with another compact object. 

\subsection{More on the Properties of Compact Objects}

\subsubsection*{Neutron Stars}

For a non-spinning NS, the spacetime outside it is described by the Schwarzschild metric, which leads to the inner most circular orbit distance or Last Stable Orbit LSO distance $d_{LSO} = 6Gm/c^2$, with $m$ the mass of the star. For the case of a spinning NS, if the orbiting object is in prograde motion, to first order in $\chi$, the LSO is $d_{LSO, \Omega} = 6Gm/c^2 (1 - 0.54433 \chi)$ \cite{Miller:1998gr}, with $\chi$ being the dimensionless spin parameter defined below. The moment of inertia for a NS, can be approximated as \cite{Pethick}

\begin{equation}
    I = 0.21 \frac{m \ell^2}{1 - 2Gm/\ell c^2},
\end{equation}

\noindent where $\ell$ is the radius of the NS. We take the speed of sound inside the star $c_s \lesssim c/\sqrt{3}$ \cite{Bedaque:2014sqa}.

For the dynamical component of the tidal effects, the dimensionless dynamical Love number $k'$, related to $n'_{g}$, is obtained from the response function  of the NS in terms of the frequency $\omega_f$, and dimensionful overlap integral, $\mathcal{I}_{f}$, of the fundamental mode of the star \cite{Chakrabarti:2013xza}, which describes to which extent an external field excites the mode.  $\mathcal{I}_{f}$ corresponds to $\mathcal{I}_{02}$ in \cite{Chakrabarti:2013xza},  where the subscript indicates the fundamental mode, $l=0$, and the  quadrupolar moment, $k=k_2$.
The response function in fourier space reads \cite{Chakrabarti:2013xza}

\begin{flalign}
\begin{split}
    \mathcal{F}(Q^{ab}) & = \frac{1}{2} \frac{\ell^5 }{G}  \frac{q_f^2}{\ell^2 (\omega_{f}^2 - \omega^2)/c^2} \mathcal{F}(E^{ab}) = \frac{1}{2} \frac{\ell^5}{G}  \frac{q_f^2}{\ell^2 \omega_f^2/c^2} \left( 1 + \frac{\omega^2}{\omega_{f}^2} + .\;.\;.\right) \mathcal{F}(E^{ab}),
\end{split}    
\end{flalign}

\noindent where we have expanded over $\omega/\omega_{f}$, and $q_{f}$, is the dimensionless overlap integral which is related to the $\mathcal{I}_f$ through $q_{f}^2 = G \mathcal{I}^2_{f} / \ell^3$. The combination, $\ell \omega_f/c$, is dimensionless as well. 

We identify the Love coefficient

\begin{equation}
    n_g = \frac{1}{2} \frac{ \ell^5}{G}  \frac{q_{f}^2}{\ell^2 \omega^2_{f}/c^2},
\end{equation}

\noindent and comparing to the Newtonian tidal Love number, $n_g =  2 k \ell^5/3G$, we identify the dimensionless Love number $k$ 

\begin{equation}
    k = \frac{3}{4} \frac{q_{f}^2}{\ell^2 \omega^2_{f}/c^2},
\end{equation}

\noindent from which we find agreement with $k$ obtained using the Clairaut-Radau equation \cite{poisson2014gravity}. From the above expansion of the response function, we can extract the term $n'_{g}$

\begin{equation}
    n'_{g} = \frac{1}{2}\frac{\ell^7}{G c^2} \frac{q_{f}^2}{\ell^4 \omega_{f}^4 /c^4} = \frac{1}{2}\frac{\ell^7}{G c^2} k' ,
\end{equation}

\noindent for a limit in which $\omega/\omega_f \ll 1$. 

For all the considered NS simulations, we consider the mass and radius defined above, even when considering different equations of state. From the considered speed of sound inside the NS, and its radius, we can safely add spin, $\chi \leq 0.2$, without breaking down the theory. 

\subsubsection*{Black Holes}

 The radius of a Schwarzschild BH is $\ell_S = 2GM/c^2$. The LSO distance for a nonspinning BH is $d_{LSO} = 6Gm/c^2$. For a spinning BH, the LSO is $d_{LSO} = 6Gm/c^2 - 2\sqrt{\frac{2}{3}} \Omega_J$ \cite{Jefremov:2015gza}, with $\Omega_J$ the angular velocity projected into the angular momentum of the BH. The moment of inertia of a BH is 
 
 \begin{equation}
    I = 4\frac{m^3G^2}{c^3}.
\end{equation}

\subsection{Slowly Spinning Objects}

As previously mentioned, our theory is valid as long as, $\Omega \ell \ll c_{s}$ holds, which can be tested once we specify the properties of the objects. First, it is important to note that instead of specifying values of the angular velocity, one works with the dimensionless spin, $\vec{\chi}$, 

\begin{equation}
    \chi = \frac{c J}{G M^2} =  \frac{c I \Omega}{G M^2},
    \label{eq:chi}
\end{equation}

\noindent where, $J = I \Omega$, is the scalar value of the angular momentum of the star.
 In order for our theory to work appropriately, we need $\chi$, to be slow $   \chi \ll 1$, to not break down the theory. Plugging in the values for a compact object we find that compact objects with the characteristic of the so far detected compact objects, will rotate slowly.

\section{Discussion}
\label{sec:discussion}

In this master's thesis part 1 of 3, we have reviewed and extended the model for spinning extended objects introduced in \cite{Delacretaz:2014oxa}, which is derived using the coset construction \cite{Callan:1969sn,Ivanov:1981wn}, a very powerful method that allows us to construct an effective theory from the symmetry breaking pattern as the only input. In this approach, a spinning extended object whose ground state breaks space-time symmetries, is coupled to a gravitational theory formulated as a gauge theory with local Poincaré symmetry and translations being non-linearly realized. We have included the internal structure \cite{Goldberger:2004jt, Goldberger:2005cd, Endlich:2015mke} and electromagnetic charge \cite{Goldberger:2005cd,Patil:2020dme}, such that we describe charged spinning extended objects, the most general extended object allowed in a theory of gravity such as general relativity with electrodynamics.

We have derived the covariant building blocks of the effective theory, to build up invariant operators and form an action. We built the underlying theory and matched the coefficients to the full known theory, to obtain the Einstein-Maxwell action in the vierbein formalism. Then, by recognizing the symmetry breaking pattern of a charged spinning extended object, we have built the leading order invariant operators that are allowed by the symmetries, to describe it as a worldline point particle with its properties and internal structure encoded in higher order corrections in the action. Such corrections take into account for the basic necessary ingredients to completely describe an extended object in the developed EFT. By matching the coefficients of the effective action from the literature, we have described charged spinning compact objects, such as BHs and NSs.

Although this effective theory for spinning extended objects \cite{Delacretaz:2014oxa} by construction is a low energy description of the dynamics, we have shown that spinning compact objects, which are described classically, fit into the description of "slowly" spinning. We have shown the equivalence of our effective theory to the ones currently used to obtain state of the art perturbative results of the binary dynamics \cite{Porto:2005ac,Levi:2015msa}, with the advantage that the covariant building blocks to construct the tower of invariant operators to all orders have been derived. Therefore, our work complements the aforementioned theories for spinning extended objects,   and lays on the foundations for a full description of the possible compact objects that can exists, including modified theories of gravity.

The most direct application of our derived action is on the PN expansion \cite{Martinez:2022vnx}, where we have shown that our theory reproduces the well known results for spinning \cite{Levi:2015msa} and charged \cite{Patil:2020dme} extended objects.  Moreover, novel results in the PN expansion have been derived on the internal structure of charged spinning compact objects \cite{Martinez:2022vnx}. The PN expansion can be used for extracting GWs, and in particular for performing numerical simulations in the late inspiral of the coalescence \cite{Martinez:2020loq}.

\acknowledgments

I.M. is very thankful to R. Penco, J. Steinhoff, T. Hinderer and H. S. Chia  for very valuable discussions, and to A. Weltman for her guidance and support. I.M. gratefully acknowledge support from the University of Cape Town Vice Chancellor's Future Leaders 2030 Awards programme which has generously funded this research, support from the South African Research Chairs Initiative of the Department of Science and Technology and the NRF, and support from the Educafin-JuventudEsGto Talentos de Exportacion programme.

\appendix

\bibliography{bib}

\begin{thebibliography}{10}

\bibitem{Goldberger:2004jt}
W.~D. Goldberger and I.~Z. Rothstein, ``{An Effective field theory of gravity
  for extended objects},'' {\em Phys.\ Rev.\ D}, vol.~73, p.~104029, 2006.

\bibitem{Goldberger:2005cd}
W.~D. Goldberger and I.~Z. Rothstein, ``{Dissipative effects in the worldline
  approach to black hole dynamics},'' {\em Phys. Rev. D}, vol.~73, p.~104030,
  2006.

\bibitem{Delacretaz:2014oxa}
L.~V. Delacrétaz, S.~Endlich, A.~Monin, R.~Penco, and F.~Riva,
  ``{(Re-)Inventing the Relativistic Wheel: Gravity, Cosets, and Spinning
  Objects},'' {\em JHEP}, vol.~11, p.~008, 2014.

\bibitem{Arkani-Hamed:2019ymq}
N.~Arkani-Hamed, Y.-t. Huang, and D.~O'Connell, ``{Kerr black holes as
  elementary particles},'' {\em JHEP}, vol.~01, p.~046, 2020.

\bibitem{Moynihan:2019bor}
N.~Moynihan, ``{Kerr-Newman from Minimal Coupling},'' {\em JHEP}, vol.~01,
  p.~014, 2020.

\bibitem{Guevara:2018wpp}
A.~Guevara, A.~Ochirov, and J.~Vines, ``{Scattering of Spinning Black Holes
  from Exponentiated Soft Factors},'' {\em JHEP}, vol.~09, p.~056, 2019.

\bibitem{Chung:2018kqs}
M.-Z. Chung, Y.-T. Huang, J.-W. Kim, and S.~Lee, ``{The simplest massive
  S-matrix: from minimal coupling to Black Holes},'' {\em JHEP}, vol.~04,
  p.~156, 2019.

\bibitem{Coleman:1969}
S.~Coleman, J.~Wess, and B.~Zumino, ``Structure of phenomenological
  lagrangians. i,'' {\em Phys. Rev.}, vol.~177, pp.~2239--2247, Jan 1969.

\bibitem{Callan:1969sn}
C.~G. Callan, Jr., S.~R. Coleman, J.~Wess, and B.~Zumino, ``{Structure of
  phenomenological Lagrangians. 2.},'' {\em Phys. Rev.}, vol.~177,
  pp.~2247--2250, 1969.

\bibitem{Volkov:1973vd}
D.~V. Volkov, ``{Phenomenological Lagrangians},'' {\em Fiz. Elem. Chast. Atom.
  Yadra}, vol.~4, pp.~3--41, 1973.

\bibitem{Ivanov:1975zq}
E.~A. Ivanov and V.~I. Ogievetsky, ``{The Inverse Higgs Phenomenon in Nonlinear
  Realizations},'' {\em Teor. Mat. Fiz.}, vol.~25, pp.~164--177, 1975.

\bibitem{Penco:2020kvy}
R.~Penco, ``{An Introduction to Effective Field Theories},'' 6 2020.

\bibitem{Goldstone:1961eq}
J.~Goldstone, ``{Field Theories with Superconductor Solutions},'' {\em Nuovo
  Cim.}, vol.~19, pp.~154--164, 1961.

\bibitem{Levi:2015msa}
M.~Levi and J.~Steinhoff, ``{Spinning gravitating objects in the effective
  field theory in the post-Newtonian scheme},'' {\em JHEP}, vol.~09, p.~219,
  2015.

\bibitem{Goldberger:2020fot}
W.~D. Goldberger, J.~Li, and I.~Z. Rothstein, ``{Non-conservative effects on
  spinning black holes from world-line effective field theory},'' {\em JHEP},
  vol.~06, p.~053, 2021.

\bibitem{Porto:2005ac}
R.~A. Porto, ``{Post-Newtonian corrections to the motion of spinning bodies in
  NRGR},'' {\em Phys. Rev. D}, vol.~73, p.~104031, 2006.

\bibitem{Levi:2014gsa}
M.~Levi and J.~Steinhoff, ``{Leading order finite size effects with spins for
  inspiralling compact binaries},'' {\em JHEP}, vol.~06, p.~059, 2015.

\bibitem{Endlich:2015mke}
S.~Endlich and R.~Penco, ``{Effective field theory approach to tidal dynamics
  of spinning astrophysical systems},'' {\em Phys.\ Rev.\ D}, vol.~93, no.~6,
  p.~064021, 2016.

\bibitem{Poisson:2004cw}
E.~Poisson, ``{Absorption of mass and angular momentum by a black hole:
  Time-domain formalisms for gravitational perturbations, and the small-hole /
  slow-motion approximation},'' {\em Phys. Rev. D}, vol.~70, p.~084044, 2004.

\bibitem{Chia:2020yla}
H.~S. Chia, ``{Tidal deformation and dissipation of rotating black holes},''
  {\em Phys. Rev. D}, vol.~104, no.~2, p.~024013, 2021.

\bibitem{Hinderer:2007mb}
T.~Hinderer, ``{Tidal Love numbers of neutron stars},'' {\em Astrophys. J.},
  vol.~677, pp.~1216--1220, 2008.

\bibitem{Yagi:2016bkt}
K.~Yagi and N.~Yunes, ``{Approximate Universal Relations for Neutron Stars and
  Quark Stars},'' {\em Phys. Rept.}, vol.~681, pp.~1--72, 2017.

\bibitem{Yagi:2016ejg}
K.~Yagi and N.~Yunes, ``{I-Love-Q Relations: From Compact Stars to Black
  Holes},'' {\em Class. Quant. Grav.}, vol.~33, no.~9, p.~095005, 2016.

\bibitem{Chakrabarti:2013xza}
S.~Chakrabarti, T.~Delsate, and J.~Steinhoff, ``{Effective action and linear
  response of compact objects in Newtonian gravity},'' {\em Phys.\ Rev.\ D},
  vol.~88, p.~084038, 2013.

\bibitem{Porto:2007qi}
R.~A. Porto, ``{Absorption effects due to spin in the worldline approach to
  black hole dynamics},'' {\em Phys. Rev. D}, vol.~77, p.~064026, 2008.

\bibitem{Coleman:1967ad}
S.~R. Coleman and J.~Mandula, ``{All Possible Symmetries of the S Matrix},''
  {\em Phys. Rev.}, vol.~159, pp.~1251--1256, 1967.

\bibitem{Maggiore:2005qv}
M.~Maggiore, {\em {A Modern introduction to quantum field theory}}.
\newblock 2005.

\bibitem{Ivanov:1981wn}
E.~A. Ivanov and J.~Niederle, ``{Gauge Formulation of Gravitation Theories. 1.
  The Poincare, De Sitter and Conformal Cases},'' {\em Phys. Rev. D}, vol.~25,
  p.~976, 1982.

\bibitem{Low:2001bw}
I.~Low and A.~V. Manohar, ``{Spontaneously broken space-time symmetries and
  Goldstone's theorem},'' {\em Phys. Rev. Lett.}, vol.~88, p.~101602, 2002.

\bibitem{Levi:2008nh}
M.~Levi, ``{Next to Leading Order gravitational Spin1-Spin2 coupling with
  Kaluza-Klein reduction},'' {\em Phys. Rev. D}, vol.~82, p.~064029, 2010.

\bibitem{Patil:2020dme}
R.~Patil, ``{EFT approach to general relativity: correction to EIH Lagrangian
  due to electromagnetic charge},'' {\em Gen. Rel. Grav.}, vol.~52, no.~9,
  p.~95, 2020.

\bibitem{HANSON1974498}
A.~Hanson and T.~Regge, ``The relativistic spherical top,'' {\em Annals of
  Physics}, vol.~87, no.~2, pp.~498--566, 1974.

\bibitem{Porto:2008jj}
R.~A. Porto and I.~Z. Rothstein, ``{Next to Leading Order Spin(1)Spin(1)
  Effects in the Motion of Inspiralling Compact Binaries},'' {\em Phys. Rev.
  D}, vol.~78, p.~044013, 2008.
\newblock [Erratum: Phys.Rev.D 81, 029905 (2010)].

\bibitem{Steinhoff:2021dsn}
J.~Steinhoff, T.~Hinderer, T.~Dietrich, and F.~Foucart, ``{Spin effects on
  neutron star fundamental-mode dynamical tides: Phenomenology and comparison
  to numerical simulations},'' {\em Phys. Rev. Res.}, vol.~3, no.~3, p.~033129,
  2021.

\bibitem{Steinhoff:2015ksa}
J.~Steinhoff, ``{Spin gauge symmetry in the action principle for classical
  relativistic particles},'' 1 2015.

\bibitem{Jordan:1986ug}
R.~Jordan, ``{Effective Field Equations for Expectation Values},'' {\em Phys.
  Rev. D}, vol.~33, pp.~444--454, 1986.

\bibitem{Binnington:2009bb}
T.~Binnington and E.~Poisson, ``{Relativistic theory of tidal Love numbers},''
  {\em Phys. Rev. D}, vol.~80, p.~084018, 2009.

\bibitem{Charalambous:2021mea}
P.~Charalambous, S.~Dubovsky, and M.~M. Ivanov, ``{On the Vanishing of Love
  Numbers for Kerr Black Holes},'' {\em JHEP}, vol.~05, p.~038, 2021.

\bibitem{Hui:2021vcv}
L.~Hui, A.~Joyce, R.~Penco, L.~Santoni, and A.~R. Solomon, ``{Ladder Symmetries
  of Black Holes: Implications for Love Numbers and No-Hair Theorems},'' 5
  2021.

\bibitem{Flanagan:2007ix}
E.~E. Flanagan and T.~Hinderer, ``{Constraining neutron star tidal Love numbers
  with gravitational wave detectors},'' {\em Phys. Rev. D}, vol.~77, p.~021502,
  2008.

\bibitem{Hut1981}
P.~{Hut}, ``{Tidal evolution in close binary systems.},'' {\em aap}, vol.~99,
  pp.~126--140, June 1981.

\bibitem{Miller:1998gr}
M.~C. Miller, F.~K. Lamb, and G.~B. Cook, ``{Effects of rapid stellar rotation
  on equation of state constraints derived from quasi-periodic brightness
  oscillations},'' {\em Astrophys. J.}, vol.~509, p.~793, 1998.

\bibitem{Pethick}
D.~G. {Ravenhall} and C.~J. {Pethick}, ``{Neutron Star Moments of Inertia},''
  {\em apj}, vol.~424, p.~846, Apr. 1994.

\bibitem{Bedaque:2014sqa}
P.~Bedaque and A.~W. Steiner, ``{Sound velocity bound and neutron stars},''
  {\em Phys. Rev. Lett.}, vol.~114, no.~3, p.~031103, 2015.

\bibitem{poisson2014gravity}
E.~Poisson and C.~Will, {\em Gravity: Newtonian, Post-Newtonian, Relativistic}.
\newblock Cambridge University Press, 2014.

\bibitem{Jefremov:2015gza}
P.~I. Jefremov, O.~Y. Tsupko, and G.~S. Bisnovatyi-Kogan, ``{Innermost stable
  circular orbits of spinning test particles in Schwarzschild and Kerr
  space-times},'' {\em Phys. Rev. D}, vol.~91, no.~12, p.~124030, 2015.

\bibitem{Martinez:2022vnx}
I.~Mart\'\i{}nez, ``{Modeling Compact Objects with Effective Field Theory:
  Dynamics of Binary Systems},'' 1 2022.
\newblock arXiv: 2201.00937.

\bibitem{Martinez:2020loq}
I.~Martinez and A.~Weltman, ``{Modeling Compact Objects with Effective Field
  Theory: The Binary Coalescence},'' 12 2020.
\newblock arXiv:2012.04140.

\end{thebibliography}
\end{document}